\documentclass[nofootinbib,     preprintnumbers, superscriptaddress]{revtex4-2}
\usepackage{amsmath,amssymb,amsbsy,amstext, amsthm, simplewick, amsfonts}

\usepackage{graphicx}
\usepackage{footmisc}
\usepackage{siunitx}
\usepackage{upgreek}
\usepackage{framed}
\usepackage{wrapfig}
\usepackage{multirow}
\usepackage{bbm}
\usepackage[dvipsnames,table]{xcolor}
\usepackage[colorlinks=true,pdfstartview=FitV,breaklinks=true]{hyperref}
\usepackage{url}
\hypersetup{urlcolor=BlueViolet,
	    citecolor=Plum,
	    linkcolor=PineGreen}

\usepackage[absolute,overlay]{textpos}

\newcommand{\equaref}[1]{Eq.~(\ref{#1})}

\newcommand{\figref}[1]{Fig.~\ref{#1}}

\newcommand{\dd}{\mathrm{d}}

\newcommand{\secref}[1]{Section~\ref{#1}}
\newcommand{\appref}[1]{Appendix~\ref{#1}}
\newcommand{\tabref}[1]{Table~\ref{#1}}

\begin{document}
\preprint{YITP-26-164}
\title{Multiscale Zoom-in Lattice Simulation for Inflation}

\author{Xinpeng Wang}
\affiliation{Kavli Institute for the Physics and Mathematics of the Universe (WPI),
The University of Tokyo Institutes for Advanced Study,
The University of Tokyo, Chiba 277-8583, Japan.}
\affiliation{Department of Physics, Graduate School of Science,
The University of Tokyo, Tokyo 113-0033, Japan.}

\author{Angelo Caravano}
\affiliation{Gravitation and Astroparticle Physics (GRAPPA),
University of Amsterdam Science Park 904, 1098 XH Amsterdam, Netherlands.}

\author{Leander Thiele}
\affiliation{Kavli Institute for the Physics and Mathematics of the Universe (WPI),
The University of Tokyo Institutes for Advanced Study,
The University of Tokyo, Chiba 277-8583, Japan.}
\affiliation{Center for Data-Driven Discovery, Kavli IPMU, UTIAS, The University of Tokyo, Kashiwa, Chiba 277-8583, Japan}

\author{Misao Sasaki}
\affiliation{Asia Pacific Center for Theoretical Physics (APCTP), Pohang 37673, Republic of Korea.}
\affiliation{Kavli Institute for the Physics and Mathematics of the Universe (WPI),
The University of Tokyo Institutes for Advanced Study,
The University of Tokyo, Chiba 277-8583, Japan.}
\affiliation{Center for Gravitational Physics and Quantum Information,
Yukawa Institute for Theoretical Physics, Kyoto University, Kyoto 606-8502, Japan.}
\affiliation{Leung Center for Cosmology and Particle Astrophysics,
National Taiwan University, Taipei 10617, Taiwan.}
\affiliation{Department of Physics, Pohang University of Science and Technology (POSTECH),
Pohang 37673, Republic of Korea.}

\begin{abstract}
% \lft{would move implementation detail to end of abstract, begin with a sentence motivating the work}\\
Lattice simulations are a powerful tool for studying the nonlinear evolution of inflationary fluctuations. The exponential expansion, however, makes it challenging to cover a broad range of scales relevant for the dynamics, limiting how many e-folds a single fixed grid can follow.
We develop a three-dimensional zoom-in lattice method to follow inflationary fluctuations over a broad range of scales from the adiabatic subhorizon regime through horizon crossing. Every few e-folds, the grid is refined into one or more subgrids, with small-scale power injected and large-scale fluctuations inherited from previous generations. Each zoom shifts the resolved band while keeping its dynamic range fixed.
This extends the duration over which lattice simulations can follow inflationary fluctuations from a few e-folds to tens of e-folds. In a Starobinsky benchmark evolved for 50 e-folds, the curvature power spectrum agrees with linear theory. In a model with an oscillatory feature, the zoom-in simulations recover the peak seen in a high-resolution fixed-grid run and show scale-dependent non-Gaussianity.

\end{abstract}

\maketitle

% \lft{i would recommend to convert the figures to PNG. with the current embedded PDFs the pages load slowly}\ac{i think that for some figures (fig 6 and 8) the pdf is better because it is higher resolution and we can zoom in to see better all the points}

\section{Introduction}
Inflation predicts that quantum fluctuations of light fields are stretched to superhorizon scales and provide the seeds for late-time curvature perturbations. In linear perturbation theory, this process is usually followed mode by mode using the Mukhanov-Sasaki equation \cite{Mukhanov:1985rz,Sasaki:1986hm}. Numerical simulations of inflation can probe effects beyond linear perturbation theory, including mode coupling and backreaction
\cite{Felder:2000hq, Figueroa:2021yhd, Caravano:2021pgc, Caravano:2025klk}. 
Inflation can be simulated using several complementary approaches. Simulations based on the stochastic formalism \cite{Starobinsky:1994bd, Mizuguchi:2024kbl,Murata:2026yqb} coarse-grain subhorizon fluctuations into a noise source, making it efficient for studying the statistics of long-wavelength perturbations and their backreaction, at the price of losing detailed spatial information and gradient effects. Numerical relativity \cite{Shibata:1995we, Baumgarte:1998te, Clough:2016ymm, Launay:2025kef} solves the Einstein-scalar system and therefore includes fully nonlinear metric backreaction, but is considerably more expensive. Lattice simulations \cite{Caravano:2021pgc,Caravano:2022epk,Caravano:2022yyv} evolve real-space field configurations in fixed spatially flat FLRW spacetime and can capture mode coupling and nonlinear scalar-field dynamics, while neglecting nonlinear dynamics in the gravitational sector.

A common challenge for these approaches is to account for the wide range of comoving scales that cross the Hubble radius during inflation. For a single periodic lattice, the range of resolved wavenumbers is set by the number of lattice points along each axis,
\begin{equation}
    \frac{k_{\rm Ny}}{k_{\rm IR}} =\frac{N_{\rm pts}}{2},
\end{equation}
where $N_{\rm pts}$ is the number of lattice points along each spatial dimension, $k_{\rm Ny}$ the Nyquist frequency and $k_{\rm IR}=2\pi/L$ the fundamental mode of a box with size $L$.

This limitation becomes severe for an inflationary evolution lasting more than $\sim 10$ e-folds. During slow roll, the comoving Hubble radius decreases approximately exponentially. A single lattice capable of resolving the full range of modes that cross the Hubble radius over 50 e-folds would therefore require
\begin{equation}
    \frac{k_{\rm Ny}}{k_{\rm IR}} \gtrsim e^{50},
\end{equation} 
corresponding to roughly $10^{22}$ lattice points along each axis.

To alleviate this limitation, we develop a zoom-in strategy that allows the inflationary evolution to be followed over a much longer interval. Rather than attempting to resolve the entire range of relevant scales on a single lattice, we track a limited band of comoving modes for several e-folds and then shift the resolved window toward larger comoving wavenumbers while keeping its dynamic range fixed. Repeating this procedure until the end of inflation ensures that each scale of interest is resolved for several e-folds before and after horizon crossing.

\begin{figure}[t]
    \centering
    \includegraphics[width=0.5\linewidth]{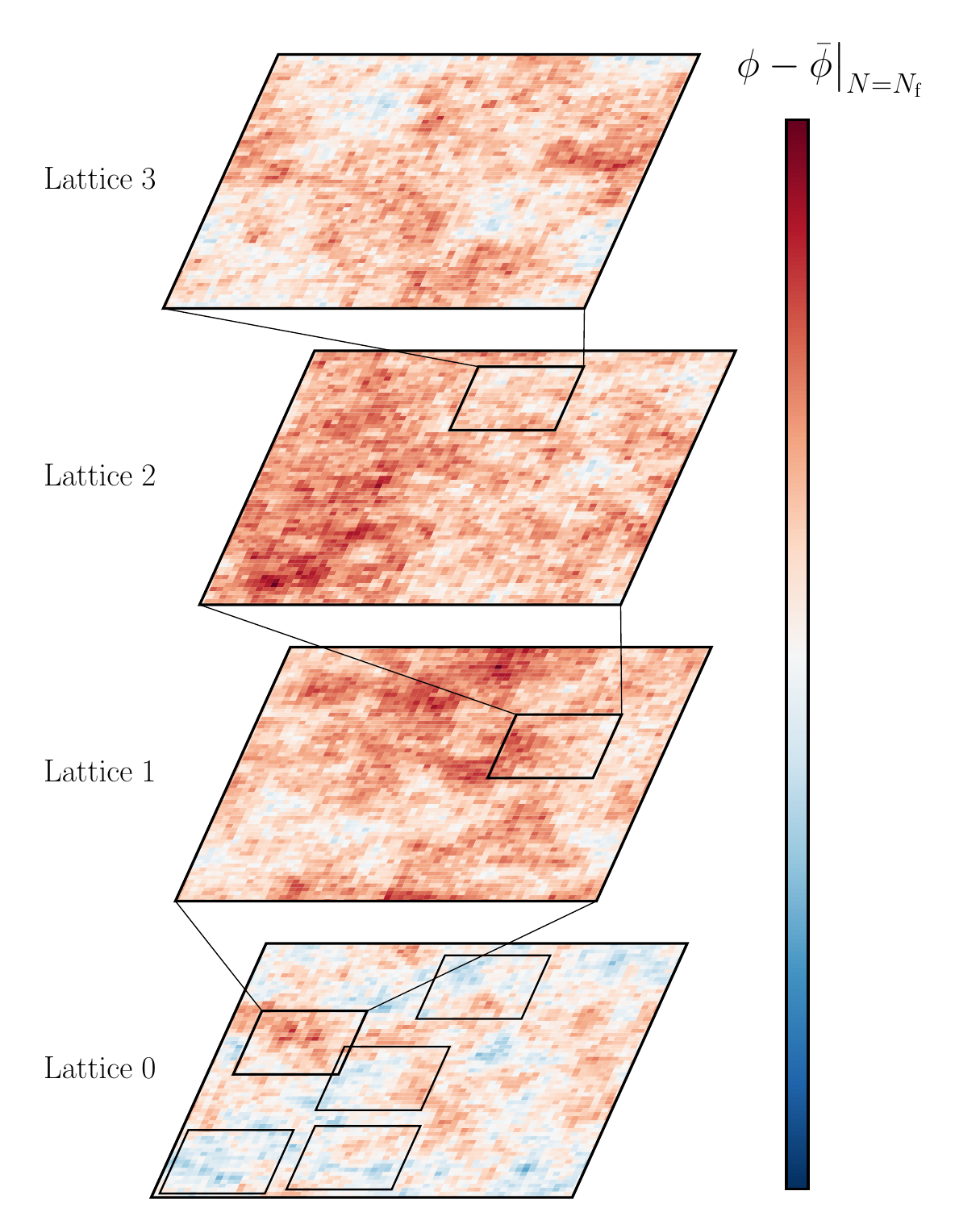}
    \caption{Schematic plot for the zoom-in scheme.}
    \label{fig:zoom_scheme}
\end{figure}

\begin{figure}[t]
    \centering   \includegraphics[width=0.5\linewidth]{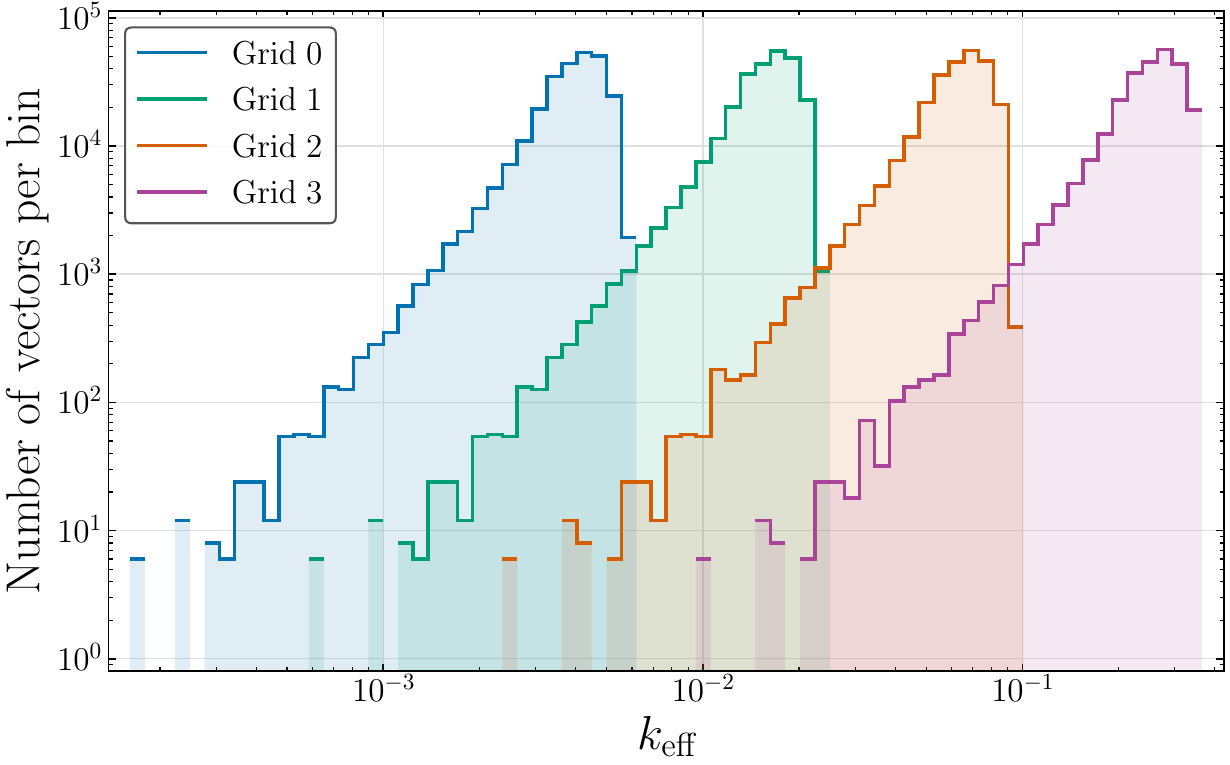}
 
    \caption{Fourier mode counts for the zoom-in scheme.}
    \label{fig:zoom_scheme_modes}
\end{figure}

In practice, the procedure begins with a parent grid that evolves a comoving volume containing long-wavelength modes. Once the physical box spans many Hubble radii, the parent configuration is refined with small-scale power injection and cropped into smaller child grids. Each child covers a smaller comoving volume while retaining the same number of lattice points, and newly resolved subhorizon modes are added to its field configuration. \figref{fig:zoom_scheme} illustrates this process. Repeating this procedure shifts the resolved lattice window toward progressively larger comoving wavenumbers, while the parent and earlier generations of grids continue to evolve the longer-wavelength modes until they become frozen.
As a result, our code extends lattice simulations of inflation to longer durations and a wider range of comoving wavenumbers without increasing the number of lattice points per grid as shown in \figref{fig:zoom_scheme_modes}.  

Because the field maps are generally not periodic in the process, we develop prescriptions for grid initialization and boundary conditions that explicitly account for this non-periodic structure. We show that, with suitable inheritance and boundary condition prescriptions, the inflaton field and its velocity on a refined patch agree closely, point by point, with those in a reference high-resolution simulation that covers the full parent volume at the same fine spatial resolution. We also reconstruct a fine-resolution field map over the full parent volume by combining the parent field with a periodic extension of the child correction.

We first benchmark the zoom-in method with Starobinsky inflation, obtaining a curvature power spectrum in good agreement with linear theory and a one-point probability density function (PDF) consistent with a Gaussian distribution. We then study a potential with an oscillatory feature and show that our zoom-in simulations recover the power spectrum seen in a high-resolution fixed-grid run. The zoom-in simulations also capture scale-dependent non-Gaussianity and an enhancement of the equilateral bispectrum near the power-spectrum peak. 

% \lft{in principle you can compute all sorts of summary statistics, so I think this is too detailed for the introduction} 

The paper is organized as follows. \secref{sec:theory} presents the theoretical framework used in the code.
\secref{sec:zoom_scheme} introduces the zoom-in scheme and compares the inheritance and boundary prescriptions. \secref{sec:statistics} describes how we estimate the curvature power spectrum and how we reconstruct the nonlinear curvature map using the $\delta N$ formalism. \secref{sec:starobinsky} and \secref{sec:oscillatory_starobinsky} present tests of Starobinsky inflation and an application to the Starobinsky potential with oscillatory features, respectively. We conclude by discussing the limitations of the method and possible extensions. Technical details of lattice dispersion, mode initialization, the non-periodic power spectrum estimator, and reconstruction of the refined field map are collected in the appendices.

Throughout this paper, we use reduced Planck units, setting
$c=\hbar=1$ and $M_{\mathrm{Pl}}\equiv(8\pi G)^{-1/2}=1$.

\section{Theory}
\label{sec:theory}
Our simulation evolves a canonical scalar field $\phi$ on a spatially flat FLRW background, including the linear gravitational correction to the scalar-field equation.
The code uses the forward number of e-folds
\begin{equation}
    N\equiv \ln a, \qquad \dd N=H\dd t=aH\dd\tau,
\end{equation}
as the time variable. The field equation is 
\begin{equation}
\partial_{NN}\phi+\left(3+\partial_N\ln H\right)\partial_N\phi
    -\frac{1}{a^2H^2}\nabla^2\phi+\frac{V_{\phi}}{H^2}+\frac{\Delta m_g^2}{H^2}(\phi-\langle\phi\rangle)=0, 
    \label{eq:eom_e-fold}
\end{equation}
where $\langle\phi\rangle$ denotes the volume-averaged field value on the lattice, and $\Delta m_g^2\equiv-2\epsilon_H H^2(3+\epsilon_H-2\eta_H)$, is the effective mass correction introduced by first-order metric perturbations in the linear scalar perturbation equation in spatially flat gauge. 
This term incorporates the gravitational effects of scalar metric perturbations on the field dynamics at linear order \cite{Caravano:2024xsb,Jamieson:2025ngu,Caravano:2025klk}. It prevents spurious superhorizon evolution, such as the failure of curvature perturbations to freeze, which can accumulate over many e-folds in long simulations \cite{Barker:2026yyg}. The Hubble slow-roll parameters are $\epsilon_H=-\dot{H}/H^2$ and $\eta_H=-\ddot{\bar\phi}/(H\dot{\bar\phi})$, where dots denote derivatives with respect to cosmic time and $\bar\phi$ is the background field.

% This term guarantees the linear perturbation evolves follows the Mukhanov-Sasaki equation strictly.
% Here and below, $\partial_N$ denotes $\partial/\partial N$. As pointed out in \cite{Caravano:2024moy, Barker:2026yyg}, the previous lattice simulations that were built on spatially flat FLRW neglects the gravitational backreactions. This omission modifies the linear mode equation
% and can lead to a disagreement between the lattice result and the linear prediction on superhorizon scales. In particular, the curvature perturbation does not freeze after horizon exit, but instead appears spurious scaling
% \begin{equation}
%     |\mathcal{R}_k(N)|
%     \propto
%     \left(\frac{H(N)}{H_\star}\right)^2 ,
% \end{equation}
% where $H_\star$ is evaluated at a time after horizon crossing when $\mathcal{R}_k(N)$ should have already been frozen. This effect would be negligible in conventional single field slow roll inflation simulated for a short period of time when Hubble parameter doesn't significantly change, however, would especially appears as an issue in simulations when grids are simulated for longer time. A originally near scale invariant power spectrum will scale as $\mathcal{P}_{\mathcal{R}}\propto(H/H_\star)^4$.

% A detailed derivation is shown in
% Appendix~\ref{app:metriccorrection}. 
The Friedmann equations are used to determine the volume-averaged Hubble parameter,
\begin{align}
     {H}^2 &=\frac{1}{3}\langle\rho\rangle ,\\
   2H^2+H\partial_NH &=\frac{1}{6}(\langle\rho\rangle-3\langle p\rangle),
\end{align}
% \xinpeng{metric perturbation appears as second-order corrections, maybe important to check from background solution.}
where the volume-averaged energy density and pressure are
\begin{align}
    \langle\rho\rangle
    &=\frac{H^2}{2}\left\langle(\partial_N\phi)^2\right\rangle
      +\frac{1}{2a^2}\left\langle|\nabla\phi|^2\right\rangle
      +\langle V\rangle,\\
    \langle p\rangle
    &=\frac{H^2}{2}\left\langle(\partial_N\phi)^2\right\rangle
      -\frac{1}{6a^2}\left\langle|\nabla\phi|^2\right\rangle
      -\langle V\rangle .
\end{align}
In practice, the code evaluates $H$ using the constraint
\begin{equation}
    H^2=
    \frac{\langle V\rangle+\left\langle|\nabla\phi|^2\right\rangle/(2a^2)}
    {3-\left\langle(\partial_N\phi)^2\right\rangle/2},
    \label{eq:bare_hubble_constraint}
\end{equation}
at every time step and $\partial_N\ln H$ is also stored as a consistency check. Since the contributions of metric perturbations to the Friedmann equations arise only at second order in field perturbations, we neglect them in the present treatment.
Before starting the nonlinear lattice evolution, we can solve the homogeneous background equations with the same initial conditions as the lattice mean field and the linear field perturbation equation in spatially flat gauge with adiabatic vacuum initial conditions. These solutions provide homogeneous and linear references for the simulation and help identify features in the inflationary trajectory that inform the choice of zoom-in settings.

Because we are interested in comparing the simulation results with linear perturbation theory, we will also solve the equation for a homogeneous scalar field $\bar\phi$
\begin{equation}
    \partial_{NN}\bar{\phi}
    +\left(3+\partial_N\ln\bar H\right)\partial_N\bar{\phi}
    +\frac{{V_{\phi}}(\bar\phi)}{\bar H^2}=0,
    \label{eq:background_eom_e-fold}
\end{equation}
together with
\begin{equation}
    \bar {H}^2
    =\frac{V(\bar\phi)}
    {3-(\partial_N\bar{\phi})^2/2},
    \qquad
    \partial_N\ln\bar H
    =-\frac{1}{2}(\partial_N\bar{\phi})^2.
\end{equation}
Linear perturbations around this background follow
\begin{equation}
\label{eq:mseq}
    \partial_{NN}\delta\phi_{k}
    +\left(3+\partial_N\ln H\right)\partial_N\delta\phi_{k}
    +\left[
        \frac{k_{\rm eff}^2}{a^2H^2}
        +\frac{V_{\phi\phi}}{H^2}
        +\frac{\Delta m_g^2}{H^2}
    \right]\delta\phi_k=0,
\end{equation}
where the initial conditions are discussed in \appref{app:initialization}. We will numerically solve this equation to compare the lattice result with linear perturbation theory.

\section{The zoom-in scheme}
\label{sec:zoom_scheme}
The purpose of the zoom-in scheme is to follow a moving band of comoving scales to capture their dynamics around horizon crossing, without resolving the entire inflationary volume at the finest lattice spacing.
The construction involves two separate choices: how the field and its velocity are inherited at each refinement, and how the environment outside the child volume is represented during its evolution. Both choices require careful handling of boundary conditions in the child grid.

\subsection{Refinement and the resolved band}
\label{sec:zoom_geometry}
We start with a parent grid, referred to as grid 0, containing $N_{\rm pts}^3$ points in a comoving box of side length $L_0$. For an integer refinement
factor $A>1$, successive generations have
\begin{equation}
    L_{i+1}=\frac{L_i}{A},\qquad
    \Delta x_{i+1}=\frac{\Delta x_i}{A},\qquad
    \Delta x_i=\frac{L_i}{N_{\rm pts}}.
    \label{eq:zoom_geometry}
\end{equation}
The fundamental and axis Nyquist wavenumbers are
\begin{equation}
    k_{\rm IR}^{i}=\frac{2\pi}{L_i},\qquad
    k_{\rm Ny}^{i}=\frac{\pi N_{\rm pts}}{L_i},\qquad
    \frac{k_{\rm Ny}^{i}}{k_{\rm IR}^{i}}=\frac{N_{\rm pts}}{2}.
    \label{eq:zoom_band}
\end{equation}
Both ends of this band shift by a factor $A$ at each zoom, while its dynamic range, set by the number of lattice points, remains fixed (see \figref{fig:zoom_scheme_modes}).
For a simulation involving multiple zooms, we define the common background field and field velocity as the volume averages on the evolving grid 0. The common Hubble parameter $\bar H$ is determined by the Friedmann constraint using the volume-averaged kinetic, potential, and gradient energies on that grid.
All grids therefore evolve using the same e-fold coordinate and time mapping, $\dd t=\dd N/\bar H$.

A zoom is triggered when the physical size of the parent box reaches a specified multiple of the Hubble radius,
\begin{equation}
    aL_i\bar H=b_{\rm pre},\qquad
    aL_{i+1}\bar H=\frac{b_{\rm pre}}{A}\equiv b_{\rm post}.
    \label{eq:zoom_trigger}
\end{equation}
For repeated zooms with the same thresholds $b_{\rm pre}$ and zoom-in factor $A$, their time separation obeys
\begin{equation}
    \int_{N_i}^{N_{i+1}}(1-\epsilon_H)\,\dd N=\ln A,
    \label{eq:zoom_interval}
\end{equation}
where $N_i$ and $N_{i+1}$ denote the times of the $i$-th and $(i+1)$-th zooms.  During slow roll, $\epsilon_H\ll1$, so the interval is approximately $N_{i+1}-N_i\simeq\ln A$. To include a certain amount of long-short mixing across different generations of grids, the choice of $b_{\rm pre}$ and $A$ must also leave an overlap between the parent and child bands, which usually requires $A\ll N_{\rm pts}/2$. Details of initializing and injecting modes during the zoom-in are explained in \appref{app:initialization}. 
At the first zoom, one or more child subvolumes are selected within the parent to continue the nonlinear evolution. We can follow a fixed number of branches from the beginning, taking only one subvolume each zoom-in after the first one, rather than generating an exponentially growing tree. 
% However, it will also be interesting to generate an exponentially growing tree for a relatively short run. For example, for a run covering a total dynamic range of 1two e-folds with $\mathrm{ln}A=4$, if we pick 8 subvolumes within 1 branch every time zooming in, we will obtain $$
\subsection{Constructing the child lattices}
\label{sec:zoom_inheritance}
Constructing the child grids is challenging mainly because of the periodicity of the box. To illustrate this, we start by assuming a \textit{periodic} parent lattice, which we call lattice 0, 
% \lft{I think you could reduce cognitive load on the reader by removing all expressions for $\partial_N\phi$ and simply saying that all expressions are entirely analogous (which they are, right?). It is good to be explicit but you also need to keep in mind that the reader has finite capacity and they will ask for every equation, is there anything unexpected written here?} 
% \lft{how about unifying the notation to the $0, \ldots, n-1, n$ convention you use below? I think $p$, $c$ are complicated especially as you change it later (and $p$ looks like a momentum at first)}

\begin{align}
    \phi_0(\mathbf x)=
    \sum_{ | q_{0}^j|< q_\mathrm{max},~ j=x,y,z}c_{0,\mathbf q_0}e^{i{2\pi\mathbf{q}_0}\cdot\mathbf x/{L_0}},\\
    % \qquad
    % \mathbf k_{p}=\frac{2\pi\mathbf{q}_p}{L_p},\qquad
    % \qquad c_{p,\mathbf{q}_p}=\frac{\widehat\phi_{p,\mathbf q_p}}{N_{\rm pts}^3},\\
        \partial_N\phi_0(\mathbf x)=
    \sum_{ | q_{0}^j|< q_\mathrm{max},~ j=x,y,z}d_{0,\mathbf q_0}e^{i{2\pi\mathbf{q}_0}\cdot\mathbf x/{L_0}},
    % \qquad
    % \mathbf k_{p}=\frac{2\pi\mathbf{q}_p}{L_p},\qquad
    % \qquad d_{p,\mathbf{q}_p}=\frac{\widehat{\partial_N\phi}_{p,\mathbf q_p}}{N_{\rm pts}^3},
    \label{eq:zoom_series}
\end{align}
where $c_{0,\mathbf q_0}$ and $d_{0,\mathbf q_0}$ are the Fourier coefficients of $\phi_0$ and $\partial_N\phi_0$, respectively. For brevity, we omit the corresponding expressions for $\partial_N\phi$ in the following, as they follow analogously from those for $\phi$. The dimensionless wavevector $\mathbf q_0$ is related to the comoving wavevector by $\mathbf k = 2\pi\mathbf q_0/L_0$, and is an integer vector because the parent box is periodic. The sums run over all modes with $|q_0^j|<q_\mathrm{max}$ for each component $j=x,y,z$. We set the cutoff to $q_\mathrm{max}=N_\mathrm{pts}/4$ to reduce discretization errors (see \appref{app:dispersion}).
% Here the hat denotes the unnormalized FFT used in Julia. 
% Embedding the same series in an $(AN_{\rm pts})^3$ Fourier array covering the same parent volume multiplies its raw FFT coefficients by $A^3$, but leaves $c_{p,\mathbf q_p}$, $d_{p,\mathbf q_p}$ and the inherited physical amplitudes unchanged. 
% This FFT normalization must not be confused with the volume compensation introduced in \secref{sec:zoom_integer}.

To construct the child lattice, one can first inject all the modes from the parent into the child box. 
%The inherited child field with higher resolution can be naively evaluated directly at the child positions.
Denoting this Fourier injection using the parent Fourier coefficients by $\mathcal I_{0\to 1}$, the field value in a child lattice at the zoom-in moment would have the form
\begin{align}
    \phi_1(\mathbf x,N)
    &=(\mathcal I_{0\to 1}\phi_0)(\mathbf x,N)
      +\delta\phi_{\rm new}(\mathbf x,N),
% \\
%     (\partial_N\phi)_1(\mathbf x,N)
%     &=(\mathcal I_{0\to 1}(\partial_N\phi)_0)(\mathbf x,N)
%       +\delta(\partial_N\phi)_{\rm new}(\mathbf x,N),
    \label{eq:zoom_cauchy}
\end{align}
where $\phi_1$ is the field on lattice 1. The first term is inherited from the parent, and $\delta\phi_{\rm new}(\mathbf x,N)$ contains the new UV modes that the parent grid does not resolve. These modes are initialized as described in \appref{app:initialization}. 
Let $\mathbf X_0$ and $\mathbf X_1$ denote the parent and child origins in the common comoving coordinate system. The parent-to-child injection $\mathcal I_{0\to 1}$ is then defined as 
\begin{align}
    (\mathcal I_{0\to 1}\phi_0)(\mathbf x,N)
    \equiv
    \sum_{|q_0^j|< q_\mathrm{max}}
  c_{0,\mathbf q_0}(N)
    e^{i(2\pi\mathbf q_0/L_0)\cdot
    (\mathbf X_1-\mathbf X_0+\mathbf x)}.
%     \\
%         (\mathcal I_{0\to 1}\partial_N\phi_0)(\mathbf x,N)
%     \equiv
%     \sum_{|q_0^j|\leq q_\mathrm{max}}
% d_{0,\mathbf q_0}(N)
%     e^{i(2\pi\mathbf q_0/L_0)\cdot
%     (\mathbf X_1-\mathbf X_0+\mathbf x)}.
\end{align}
% Equivalently, the parent Fourier components are inherited with
% wavevectors $\mathbf q_c=\mathbf q_p/A$ defined on child grid. 
On the child grid, the same comoving wavevector is labeled by $\mathbf q_1$, with $\mathbf k = 2\pi\mathbf q_1/L_1$. Since $L_1 = L_0/A$, this gives $\mathbf q_1 = \mathbf q_0/A$. One should notice that the child-grid wavevector $\mathbf q_1$ is not guaranteed to be an integer vector in general. In other words, even when the parent is periodic, a cropped child patch is generally not periodic over $L_1=L_0/A$. We refer to the Fourier modes of the non-periodic components as fractional modes with respect to the child box. 
% A parent mode, expressed in coordinates relative to the child origin $\mathbf X_c$, becomes
% \begin{equation}
%     e^{i\mathbf k_{p}\cdot(\mathbf X_c+\mathbf x)}
%     =e^{i\mathbf k_{p}\cdot\mathbf X_c}
%      e^{2\pi i(\mathbf q_p/A)\cdot\mathbf x/L_c}.
% \end{equation}
% It belongs to the child's periodic Fourier basis only if $\mathbf{q}_p/A\in \mathbb Z^3$.

\begin{figure}[t]
    \centering
    \includegraphics[width=1\linewidth]{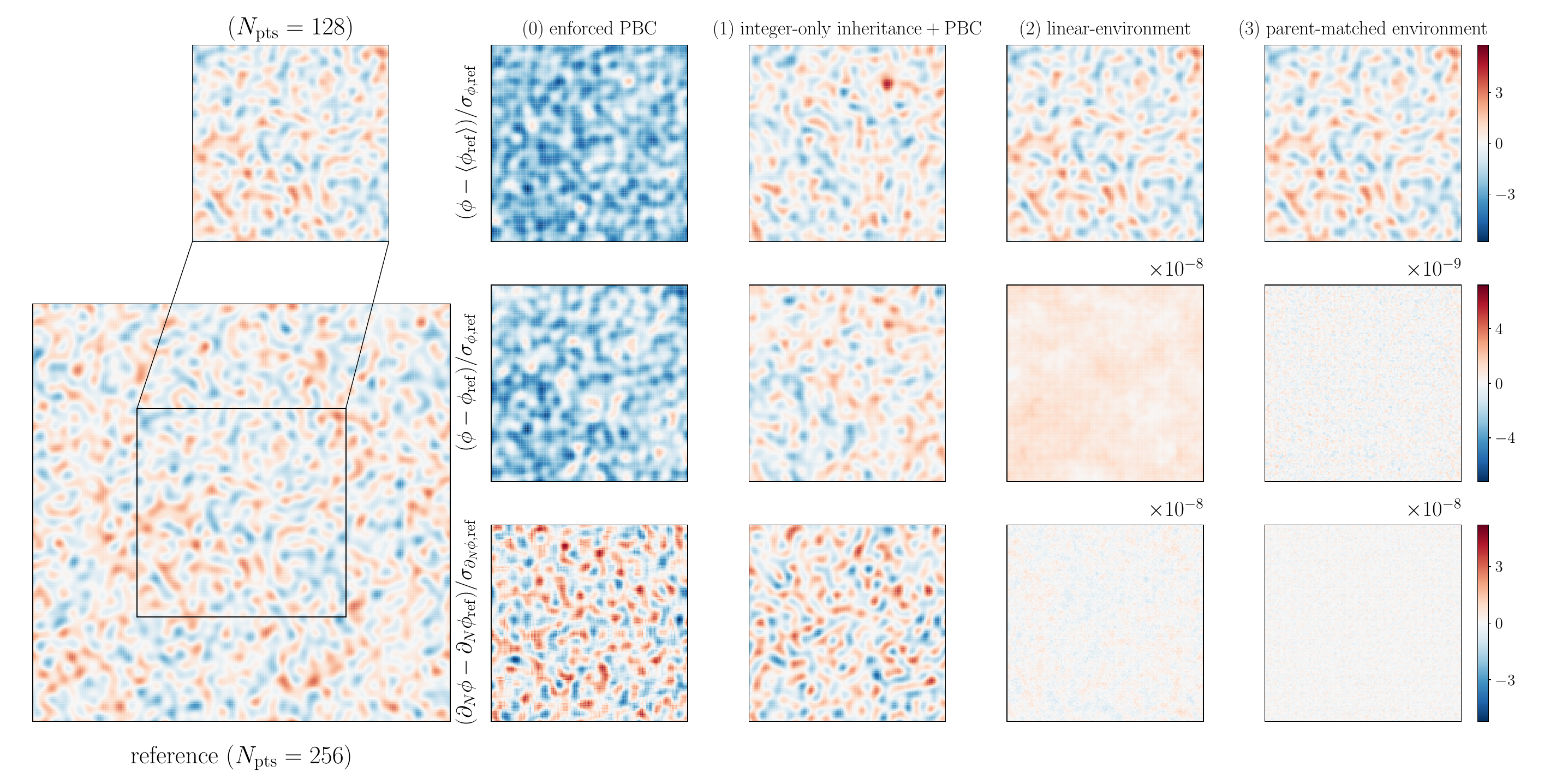}
    \caption{Comparison of field maps after two e-folds of evolution in the Starobinsky potential. The $256^3$ reference simulation covers the same comoving volume as the $128^3$ parent grid, with twice as many lattice points along each axis. The maps on the left show a two-dimensional slice of the full reference and the corresponding child-sized crop, both at the final time. Each child contains $128^3$ lattice points and has the same lattice spacing as the reference simulation. The remaining columns show (0) full field inheritance with enforced periodic boundary conditions (PBC), (1) power-normalized integer inheritance with PBC, (2) a linearly evolved environment, and (3) an environment supplied by the nonlinearly evolved parent.  Cases (0), (2), and (3) start from the same total field $\phi$ and field velocity $\partial_N \phi$ as the reference within the child region at initialization, whereas case (1) modifies both through integer-only reconstruction.}
   \label{fig:zoom_boundary_comparison}
\end{figure}

Handling these additional fractional modes is nontrivial for two main reasons. First, these non-periodic modes require a special boundary condition prescription during the evolution. Second, these fractional modes distribute their power across the periodic child modes when performing FFT, an effect known as spectral leakage \cite{Harris:1978yaq}. Naively, one could simply ignore this issue and enforce periodic boundary conditions (PBCs) on the child. However, due to the presence of these fractional modes inherited from the parent, this would introduce major artifacts in the simulation, as illustrated in \figref{fig:zoom_boundary_comparison}.
 % And, just ignoring the long-wavelength background is apparently incorrect. 
% A wave can have a wavelength shorter than the child box size $L_c$ and be well sampled by its cells, while still not satisfying $\mathbf{q}_p/A\in \mathbb Z^3$. 
% The FFT of a cropped fractional wave distributes its power across several child Fourier coefficients, an effect known as spectral leakage.
% Also, applying periodic boundary conditions (PBC) to a child box doesn't reflect the parent field's configuration. 

A second difficulty arises because the parent itself need not be periodic in the first place, as it could itself be a child from a previous zoom-in. Therefore, we have to handle two kinds of fractional modes: those that first become fractional at the current zoom-in and those that were already fractional on the parent grid. To do that, in the following we will consider three ways of treating the boundary conditions for these fractional modes. The first changes the inherited field so that it is periodic in the child box by construction. The other two preserve the inherited field in real space and represent its non-periodic environment explicitly. \figref{fig:zoom_boundary_comparison} illustrates the three methods alongside the enforced PBC result.

\subsubsection{Sparse integer inheritance with periodic boundaries}
\label{sec:zoom_integer}

In the first prescription, only modes that are periodic in the child grids are retained and the fractional modes are dropped.  But we compensate for the power lost by this procedure with an increased power assigned to each periodic mode. Considering the parent lattice labeled by $n-1$ and child lattice labeled by $n$, we have
\begin{equation}
    L_{n}^3|c_{n,\mathbf q_{n}}|^2
    =L_{n-1}^3|c_{n-1,A\mathbf q_{n}}|^2,
    \label{eq:zoom_sparse_power}
\end{equation}
where $\mathbf q_n=A^{-1}\mathbf q_{n-1}$ is an integer vector.  Therefore, except for the zero mode, the inherited Fourier coefficients on the child lattice satisfy 
\begin{align}
 %   c_{c,\mathbf q}
 % &=A^{3/2}
 % e^{i\mathbf k_{p,A\mathbf q}\cdot\mathbf (X_c-\mathbf X_p)}
 % c_{p,A\mathbf q},\\
 %   d_{c,\mathbf q}
 %    &=A^{3/2}e^{i\mathbf k_{p,A\mathbf{q} _p}\cdot\mathbf (X_c-\mathbf X_p)}
 %      d_{p,A\mathbf q_p}.
    |c_{n,\mathbf q_n}|
 &=A^{3/2}
 |c_{n-1,\mathbf q_{n-1}}|.
 % \\
 %   |d_{n,\mathbf q_n}|
 %    &=A^{3/2}|d_{n-1,\mathbf q_{n-1}}|.
    \label{eq:zoom_sparse_coefficients}
\end{align}
% With our Fourier-series convention, the finite-volume estimate of the field power at a given wavevector is $L^3|c_{\mathbf q}|^2$.
% We therefore rescale the retained coefficients by
% $A^{3/2}=(L_p/L_c)^{3/2}$ so that the parent and child assign the same power to each retained comoving wavevector:
% This matching applies only to the retained modes and does not impose
% conservation of the total field variance or energy across refinement.
The child grid is therefore initialized at the zoom time as
\begin{align}
    \phi_n(\mathbf x,N)
    &=\langle\phi_{n-1}\rangle+
  A^{3/2}[(\mathcal I^I_{n-1\to n}
    \phi_{n-1})(\mathbf x,N)-\langle\phi_{n-1}\rangle]
    +\delta\phi_{\rm new}(\mathbf x,N),
% \\
%     (\partial_N\phi)_{n}(\mathbf x,N)
%     &=\langle\partial_N\phi_{n-1}\rangle+
%    A^{3/2}[(\mathcal I^I_{n-1\to n}
%     \partial_N\phi_{n-1})(\mathbf x,N)-\langle\partial_N\phi_{n-1}\rangle]+\delta(\partial_N\phi)_{\rm new}(\mathbf x,N),
%     \label{eq:zoom_sparse_integer}
\end{align}
where
\begin{align}
    (\mathcal I^I_{n-1\to n}\phi_{n-1})(\mathbf x,N)
    =
    \sum_{\mathbf q_{n-1}\in A\mathbb Z^3 }
  c_{n-1,\mathbf q_{n-1}}(N)
    e^{i(2\pi\mathbf q_{n-1}/L_{n-1})\cdot
    (\mathbf X_n-\mathbf X_{n-1}+\mathbf x)}.
  %   \\
  %   (\mathcal I^I_{n-1\to n}\partial_N\phi_{n-1})(\mathbf x,N)
  %   =
  %   \sum_{\mathbf q_{n-1}\in A\mathbb Z^3 }
  % d_{n-1,\mathbf q_{n-1}}(N)
  %   e^{i(2\pi\mathbf q_{n-1}/L_{n-1})\cdot
  %   (\mathbf X_n-\mathbf X_{n-1}+\mathbf x)}.
\end{align} 
This inverse FFT then gives a periodic configuration, which enables the grid to evolve independently of the older grids. In this case, the simple PBC and FFT are applicable within every box. This prescription drops most parent modes and changes the amplitudes of those retained. It is therefore a statistical replacement of the patch that keeps the power spectrum, but does not preserve the real space field map and higher-order correlations.

\begin{figure}[t]
    \centering
    \includegraphics[width=0.8\linewidth]{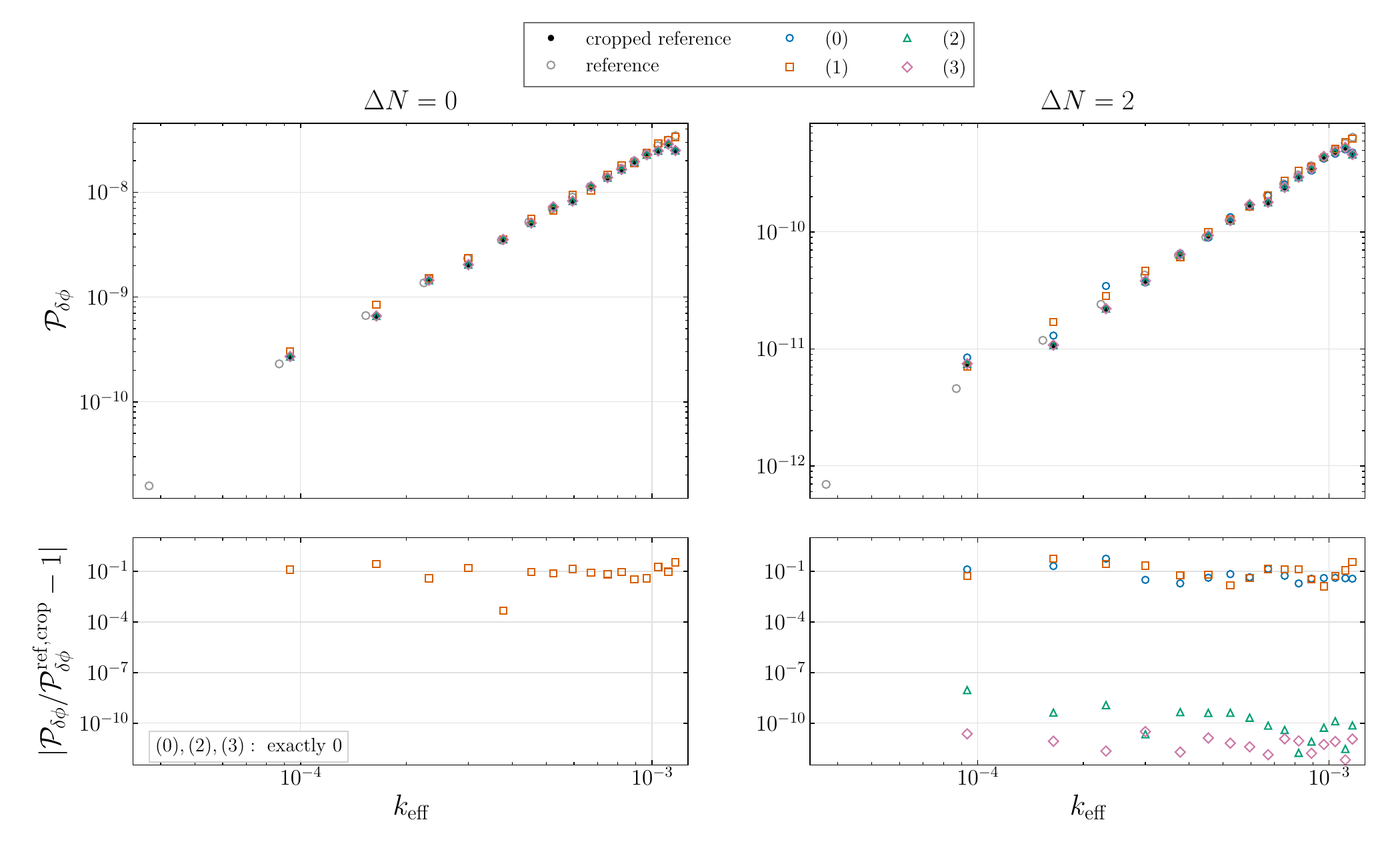}
    \caption{Power spectra obtained from FFT of field perturbations for the configurations compared in \figref{fig:zoom_boundary_comparison}. The upper panels show $\mathcal P_{\delta\phi}$ at initialization ($\Delta N=0$) and after two e-folds ($\Delta N=2$), including the full reference, its child-sized crop, and prescriptions (0)-(3). The lower panels show the absolute fractional deviations from the child-sized crop,
$\left|\mathcal P_{\delta\phi}/
\mathcal P_{\delta\phi}^{\mathrm{ref,crop}}-1\right|$.
 }
   \label{fig:zoom_boundary_comparison_powerspectrum}
\end{figure}

\subsubsection{Linear-environment method}
\label{sec:zoom_linear_environment}

Let us consider multiple zoom-ins, generating multiple generations of grids. We focus on the $n$-th generation and the previous ($0$-th to $n-1$-th) generations.  
At a zoom-in moment, when the parent lattice ($n-1$-th) already contains \textit{non-periodic} components coming from the previous generations, only its periodic components can be injected to the child lattice ($n$-th). The non-periodic components should be evaluated directly at the positions of the child grid. This can be done by performing the inverse Fourier transformation of the Fourier coefficients obtained from the previous generations of grids where those waves are still periodic. 

Let $\phi_s$ denote the field value on the $s$-th lattice ($0\leq s\leq n-1$) used to construct the inherited field on the child ($n$-th) lattice. In the general case, the child field is initialized as
\begin{align}
    \phi_n(\mathbf x,N)
    &=
    \sum_{s=0}^{n-1}(\mathcal I^\mathcal{Q}_{s\to n}
    \phi_s)(\mathbf x,N)+(\mathcal I^I_{n-1\to n}
    \phi_{n-1})(\mathbf x,N)
    +\delta\phi_{\rm new}(\mathbf x,N),
% \\
%     (\partial_N\phi)_n(\mathbf x,N)
%     &=
%     \sum_{s=0}^{n-1}(\mathcal I^\mathcal{Q}_{s\to n}
%     \partial_N\phi_s)(\mathbf x,N)
%     +(\mathcal I^I_{n-1\to n}
%     \partial_N\phi_{n-1})(\mathbf x,N)+\delta(\partial_N\phi)_{\rm new}(\mathbf x,N),
    \label{eq:field_linear_environment}
\end{align}
where the operator $\mathcal{I}^I_{n-1\to n}$ retains only the integer Fourier modes of the parent lattice that remain periodic in the child grid and $\mathcal{I}^{\mathcal{Q}}_{s\to n}$ carries all the fractional modes from previous generations.
In Eq.~\eqref{eq:field_linear_environment}, each component is expressed in the periodic Fourier basis of its source grid $s$, and evaluated at each child position according to
% \begin{equation}
%     F_s^{(n)}(\mathbf x,N)=
%     \sum_{\mathbf q_s\in \mathcal Q_s}
%     c_{s,\mathbf q_s}^{(n)}(N)
%     e^{i(2\pi\mathbf q_s/L_s)\cdot\mathbf x},
% \end{equation}
\begin{align}
    (\mathcal I^\mathcal{Q}_{s\to n}\phi_s)(\mathbf x,N)
    \equiv
    \sum_{\mathbf q_s\in\mathcal Q_s}
  c_{s,\mathbf q_s}(N)
    e^{i(2\pi\mathbf q_s/L_s)\cdot
    (\mathbf X_n-\mathbf X_s+\mathbf x)},
  %   \\
  %   (\mathcal I^\mathcal{Q}_{s\to n}\partial_N\phi_s)(\mathbf x,N)
  %   \equiv
  %   \sum_{\mathbf q_s\in\mathcal Q_s}
  % d_{s,\mathbf q_s}(N)
  %   e^{i(2\pi\mathbf q_s/L_s)\cdot
  %   (\mathbf X_n-\mathbf X_s+\mathbf x)}.
\end{align}
Here $\mathcal Q_s$ contains all of the $\mathbf{q}_s$ that first become non-integer modes when transported to the next generation, i.e.
\begin{align}
    \mathbf q_{s+1}=\frac{\mathbf q_s}{A_s}\notin\mathbb Z^3, \qquad A_s\equiv\frac{L_s}{L_{s+1}}.
\end{align}

\begin{figure}[t]
    \centering
    \includegraphics[width=1\linewidth]{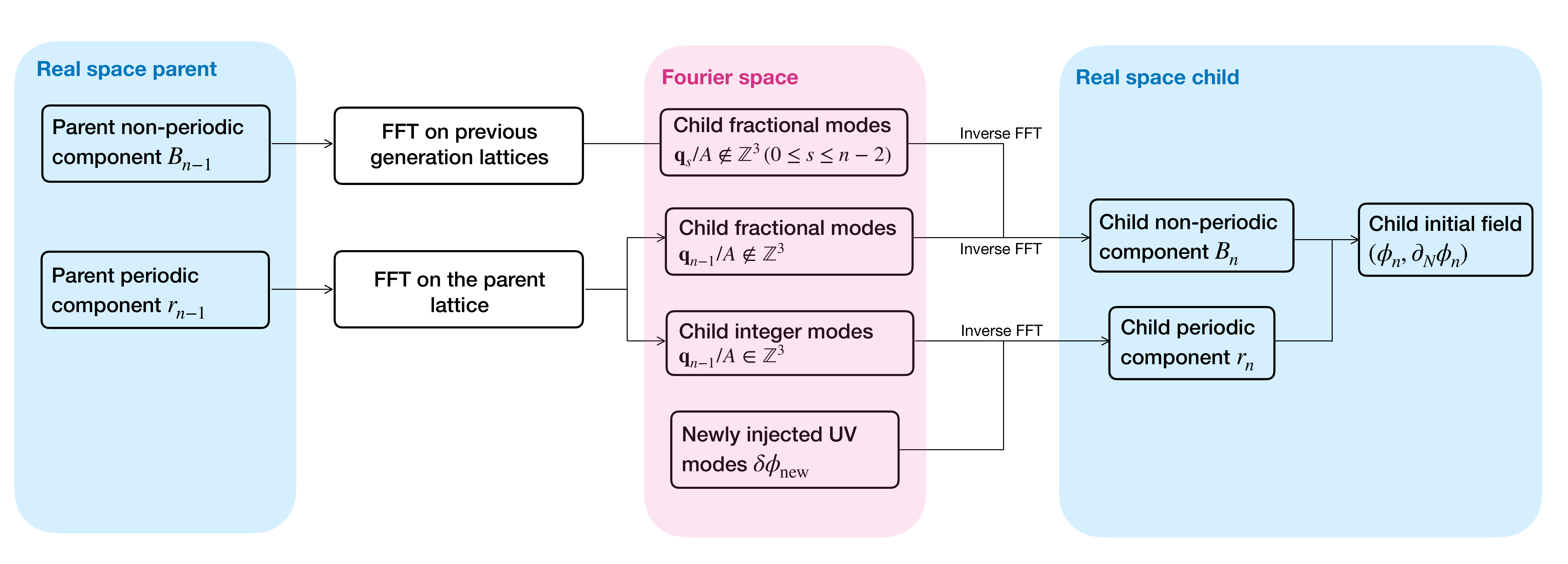}
 
    \caption{Pipeline for constructing child lattices.}
    \label{fig:pipeline}
\end{figure}

To simplify the notation, we split the field $\phi_n$ into a non-periodic component $B_n$ and a periodic component $r_n$:
\begin{equation}
    \phi_n (\mathbf{x}, N)=B_n(\mathbf{x}, N)+r_n(\mathbf{x}, N),
    \label{eq:zoom_environment_split}
\end{equation}
where $B_n$ contains the inherited contributions that are non-periodic in the child box,
\begin{align}
    B_n (\mathbf{x}, N)&=\sum_{s=0}^{n-1}(\mathcal I^\mathcal{Q}_{s\to n}
    \phi_s)(\mathbf x,N),
    \label{eq:non-periodic_env}
    \end{align}
and periodic inherited and newly injected modes remain in $r_n$. The child box construction pipeline is illustrated in \figref{fig:pipeline}.

To obtain the boundary condition, $r_n=\phi_n-B_n$ can be legitimately periodically extended. For instance, assuming we have $N_{\mathrm{pts}}$ cells along the $x$ direction, the value of the scalar field in the cell next to the boundary should be given by 
\begin{equation}
    \phi(N_{\rm pts}+1,j,k)
    =({\phi(1,j,k)-B(1,j,k)})
      +B(N_{\rm pts}+1,j,k),
    \label{eq:zoom_boundary_environment}
\end{equation}
and similarly for $\partial_N\phi$, where ${\phi(1,j,k)-B(1,j,k)}={r(1,j,k)}$ comes from the periodic component, and $B(N_{\rm pts}+1,j,k)$ is the actual value at the $N_{\rm{pts}}+1$ cell computed from the inverse Fourier transformation of the fractional modes. We note that we evolve $\phi_n$ on the $n$-th lattice following \equaref{eq:eom_e-fold}, so we do not need separation into $B$ and $r$ except for the boundary values.
% Constructing $B$ at each time step requires its Fourier coefficients $(c_{\mathbf q},d_{\mathbf q})$.

In the linear-environment method in this section, we split $\phi$ into periodic/non-periodic components, and  the non-periodic part at the boundary is evaluated by evolving the fractional modes using linear perturbation theory. To be explicit, the Fourier coefficients used to construct $B$ on the boundary are propagated from their values at the zoom-in time by the linear transfer matrix,
\begin{equation}
    \begin{pmatrix}
        c_{\mathbf q}(N+\Delta N)\\ d_{\mathbf q}(N+\Delta N)
    \end{pmatrix}
    =
    \mathbf{T}_{\mathbf q}(N+\Delta N,N)
        \begin{pmatrix}
        c_{\mathbf q}(N)\\ d_{\mathbf q}(N)
    \end{pmatrix},
    \label{eq:zoom_environment_transfer}
\end{equation}
obtained from \equaref{eq:mseq}. The effective wavenumber involved in the transfer matrix is evaluated with the spacing of the child lattice. 
% At each time step, the $B$ values on the boundary cells are provided by inverse Fourier transformation of the environmental modes. 
% In the code, for each previous lattice $s$, the field and velocity coefficients entering the $B_n$ are initialized from the Fourier coefficients of $r_s=\phi_s-B_s$ and $\partial_N r_s$ at the zoom-in moment when those modes first become non-periodic in the child box.
% At subsequent zoom-ins, previously stored components are evolved linearly together with the evolution of the grid $n$.
This method neglects future nonlinear evolution of the environment's effect on the boundaries of child grids, with a computational advantage that the parent lattices are no longer required to be evolved nonlinearly all the time, even though boundary construction is still required.

\subsubsection{Parent-matched environment method}
\label{sec:zoom_parent_environment}

The third method uses the same initial field configuration after zoom-in as the second method \equaref{eq:field_linear_environment}, and also uses the same field decomposition \equaref{eq:zoom_environment_split} and boundary construction formula \equaref{eq:zoom_boundary_environment}, but updates the environment from the nonlinearly evolved parent grids that evolve alongside the child grids.

That requires the parent grids to remain active after each zoom and continue to evolve according to \equaref{eq:eom_e-fold}. Unlike the linear-environment prescription, the boundary values are obtained by interpolating the parent fields rather than propagated by a linear transfer matrix. The nonlinear evolution resolved by the parent grids can therefore influence the child evolution.

At each time step, the boundary values are determined by the parent fields at that same stage supplied using \equaref{eq:zoom_boundary_environment}, where for the values of $B$ on the boundary, 
% for periodic parent field, the field at boundary positions of the child lattice can be evaluated by Fourier injection as explained in  \equaref{eq:zoom_cauchy}.
we reconstruct them using the Fourier injection of the non-periodic components \equaref{eq:non-periodic_env} with all of the Fourier coefficients obtained from the previous lattices. 

% For a comparison in \figref{fig:zoom_boundary_comparison}, we use the  of the parent grid and the parent Laplacian is additionally matched to the child grid dispersion relation on the inherited field. That is, the parent field is evolved using a modified equation of motion whose gradient operator matches the child grid dispersion relation on the inherited Fourier band.

This method retains nonlinear environmental evolution on the scales resolved by the previous lattices. However, it requires the relevant lattices to remain active together with the child grids, making it more computationally expensive for long runs.

\subsection{Comparison and discussion}
\label{sec:comparison}
We compare the three inheritance and boundary prescriptions with an enforced PBC run for two e-folds of evolution in the Starobinsky potential, as shown in \figref{fig:zoom_boundary_comparison}.
The setup consists of a parent grid with
$N_{\mathrm{pts}}=128$ and a fine reference grid with
$N_{\mathrm{pts}}=256$, covering the same comoving volume
and sharing the same initial realization of their common
Fourier modes. The reference is evolved on the full volume,
and the corresponding child-sized patch ($N_{\mathrm{pts}}=128$) is extracted at the final time for comparison which we call the cropped reference.

The four child simulations, each with $N_{\mathrm{pts}}=128$,
use the following prescriptions:
(0) enforced PBC;
(1) sparse integer-mode inheritance with PBC;
(2) linear-environment boundary; and
(3) parent-matched boundary.
Cases (0), (2), and (3) start from identical field $\phi$ and field velocity $\partial_N \phi$ configurations, obtained by directly cropping the initial fine reference.
Case (1) uses the same underlying realization, but its initial configuration is modified by the integer-mode selection and amplitude rescaling.

The top, middle, and bottom rows show two-dimensional
slices of
\begin{equation}
    \frac{\phi-\langle\phi_{\mathrm{ref}}\rangle}
         {\sigma_{\phi,\mathrm{ref}}},
    \qquad
    \frac{\phi-\phi_{\mathrm{ref}}}
         {\sigma_{\phi,\mathrm{ref}}},
    \qquad
    \frac{\partial_N\phi-\partial_N\phi_{\mathrm{ref}}}
         {\sigma_{\partial_N\phi,\mathrm{ref}}},
\end{equation}
respectively. Here $\phi_{\mathrm{ref}}$ denotes the field in the cropped reference, and the mean $\langle\phi_{\mathrm{ref}}\rangle$ and standard deviations $\sigma_{\phi,\mathrm{ref}}$ are measured over the cropped reference at the final time $\Delta N=2$. 

The comparison highlights the importance of consistently retaining the inherited boundary values when evolving the child lattice. Starting from the same cropped field and velocity, enforcing PBCs in (0) produces a displacement of the mean field relative to the cropped reference. The final field map difference has an RMS of approximately $2.62\,\sigma_{\phi,\mathrm{ref}}$. The sparse inheritance prescription (1) also differs substantially from the reference, with an RMS difference of approximately $1.15\,\sigma_{\phi,\mathrm{ref}}$.
In this case, however, part of the discrepancy already appears at initialization because the reconstruction changes the inherited realization. Despite their different origins, both discrepancies lead to power-spectrum differences typically of order $10^{-1}$ (see \figref{fig:zoom_boundary_comparison_powerspectrum}).

By contrast, the linear-environment and parent-matched methods (2) and (3) well reproduce both the reference field configuration and its power spectrum. Their final field-map errors are approximately $10^{-8}$ and $10^{-10}$ in units of $\sigma_{\phi,\mathrm{ref}}$, respectively. As a further consistency check, we use the $128^3$ parent and the $128^3$ child to reconstruct a large-scale $256^3$ box with the same comoving volume as the parent and compare it to the reference $256^3$ simulation. We show this comparison in \appref{app:total_map_reconstruction}. 

Before proceeding to the next section, we introduce a cheaper alternative real-space method that can be used when multiple zoom-ins make the methods (2) and (3) numerically expensive. To reduce the numerical cost of Fourier transforming the parent modes onto the child grid, we instead initialize the child grid by interpolating the parent grid in real space
\begin{align}
\phi_n(\mathbf x,N)=(I_{n-1\rightarrow{n}}\phi_{n-1})(\mathbf x,N)+\delta\phi_{\mathrm{new}}(\mathbf x,N),
\end{align}
where $I_{n-1\rightarrow n}$ denotes the spatial interpolation operator from the parent lattice to the child lattice, and the new UV modes $\delta\phi_{\mathrm{new}}(\mathbf x,N)$ are initialized in Fourier space as before. The real-space interpolation uses a seventh-degree Lagrange interpolation, and provides an approximate reconstruction of the non-periodic inherited field. In this method, we also construct the child boundary values using local interpolation of the parent's real-space field. %The interpolated parent field supplies the non-periodic boundary variation, while the child periodic component relative to this field is extended periodically. 
This alternative method recovers the reference values for the field and velocity with RMS differences of approximately $1.36\%$ and $1.16\%$ for $\sigma_{\phi,\mathrm{ref}}$ and $\sigma_{\partial_N\phi,\mathrm{ref}}$, respectively. It therefore provides a cheaper alternative to methods (2) and (3) while still giving acceptable results.

\section{Statistics for the curvature perturbation}
\label{sec:statistics}

\subsection{The power spectrum}
We estimate the curvature power spectrum from the discrete Fourier transform of the lattice field after subtracting the mean field value. For a cubic box with $N_{\rm pts}^3$ points and comoving size $L$, we define the normalized Fourier coefficients by
\begin{equation}
    {\delta X}_{\mathbf k}
    =
    \frac{1}{N_{\rm pts}^3}
    \sum_{\mathbf x}
    \delta X_{\mathbf x}\,
    e^{-i\mathbf k\cdot\mathbf x}.
\end{equation}
The corresponding finite-volume power spectrum is
\begin{equation}
    P_X(k)
    =
    L^3
    \left\langle
        \left|{\delta X}_{\mathbf k}\right|^2
    \right\rangle_{\rm shell},
\end{equation}
where the average is performed over the Fourier modes in the given $k$ shell in Fourier space. In the spectrum analysis, we use the effective lattice wavenumber $ k_{\rm eff}$ associated with the finite-difference Laplacian employed in the field evolution (see \appref{app:dispersion}).
For a Fourier mode with comoving wavevector $\mathbf k$, the gradient contribution to its discrete evolution is determined by $k_{\rm eff}^2(\mathbf k)$ rather than $|\mathbf k|^2$ \cite{Caravano:2021pgc}. For the standard second-order centered Laplacian,
\begin{equation}
    k_{\rm eff}^2(\mathbf k)
    =
    \sum_{j=1}^{3}
    \left[
        \frac{2}{\Delta x}
        \sin\left(\frac{k_j\Delta x}{2}\right)
    \right]^2,
\end{equation}
where $\Delta x=L/N_{\rm pts}$ is the comoving lattice spacing.

In the calculation of the power spectrum, we use the linear relation between the comoving curvature perturbation and the field perturbation in spatially flat gauge
\begin{equation}
\mathcal{R}_{\mathbf{k}}
=
-\frac{\delta\phi_{\mathbf{k}}}
{\partial_N\bar{\phi}}.
\end{equation}
Therefore, the dimensionless curvature power spectrum in a given $k$ shell is given by
\begin{equation}
    {\mathcal{P}}_{\mathcal R}
    =
    \frac{{\mathcal P}_{\delta\phi}}
         {(\partial_N\bar\phi)^2}
    =
    \frac{V}{2\pi^2
                   (\partial_N\bar\phi)^2}
    \left\langle
        k_{\mathrm{eff}}^3(\mathbf k)
        \left|{\delta\phi}_{\mathbf k}\right|^2
    \right\rangle_{\mathrm{shell}}.
\end{equation}
For multi-child simulations, we first evaluate the spectrum separately on each child lattice in the same generation and then average the results. However, directly Fourier transforming a child field generally introduces crop-induced spectral leakage as discussed in the previous sections due to the existence of fractional modes. %i.e.\ they  project power onto several child Fourier coefficients, mixing different scales.
We therefore construct a power spectrum estimator that correctly takes into account the inherited fractional modes when computing the power associated to each bin in the simulation. This procedure is detailed in \appref{app:estimator}.

\subsection{$\delta N$ formalism and one-point statistics}
\label{sec:deltaN}
To compute the one-point statistics of the curvature perturbation, we reconstruct its nonlinear map by applying the $\delta N$ formalism \cite{Starobinsky:1985ibc,Salopek:1990jq,Sasaki:1995aw,Lyth:2004gb} to the lattice output, as in \cite{Caravano:2025diq}. At leading order in the gradient expansion, each region evolves as a separate homogeneous universe \cite{Wands:2000dp}. Starting from the lattice field and its velocity on a common spatially flat slice, we evolve each local trajectory to the same final energy density. The difference in integrated expansion relative to the homogeneous reference gives $\zeta=\delta N$, which we identify with $\mathcal R$ on superhorizon scales in the adiabatic regime considered here. We therefore write
\begin{equation}
\mathcal{R}(\mathbf{x})=\delta N(\mathbf{x})
=
N\!\left(\phi_*(\mathbf{x}),\partial_N\phi_{*}(\mathbf{x});\rho_{\rm f}\right)
-N\!\left(\bar{\phi}_*,{\partial_N\bar\phi_{*}};\rho_{\rm f}\right),
\label{eq:deltaN}
\end{equation}
where $\rho_{\rm f}=\frac{3V(\phi_\mathrm{f})}{3-1/2(\partial_N\phi_{\rm f})^2}$ is the density on the chosen final uniform-density slice. Here $N$ is the number of e-folds for a homogeneous trajectory initialized at $(\phi_*,\partial_N\phi_*)$ to reach $\rho_{\rm f}$.

To use the $\delta N$ formalism safely, we first evolve the system until the modes of interest are sufficiently superhorizon when the gradient energy is subdominant and the trajectory is in the attractor regime. At this stage, the evolution of each point follows the background dynamics and can therefore be treated as a separate universe. We then choose a common final uniform-density slice and evaluate the local expansion difference according to ~\equaref{eq:deltaN}.

% The $\delta N$ formalism does not capture the residual superhorizon evolution sourced by spatial gradient terms. In addition to requiring the modes of interest to be sufficiently superhorizon, we choose the initial slice for the $\delta N$ calculation such that no sharp feature in the
% inflationary potential occurs between this slice and the final uniform-density slice. In particular, the background evolution is required to remain on the slow-roll attractor throughout this interval, so that no additional non-slow-roll evolution is generated after the initial slice.

% To compute the one-point statistics, we first coarse-grain the lattice fields by using a normalized Gaussian window function, i.e.
% \begin{align}
%     \phi_R(\mathbf{x})=\int d^3y~G_{R}(\mathbf{x}-\mathbf{y})\phi(\mathbf{y}),\qquad G_{R}(\mathbf{r})=\frac{e^{-r^2/(2R^2)}}{(2\pi R^2)^{3/2}}.
% \end{align}

% The coarse-grained fields are then used to construct the $\mathcal{R}(\mathbf{x};R)$ through \eqref{eq:deltaN}.

To compute the one-point statistics, we first construct $\mathcal{R}(\mathbf{x})$ at each lattice point on the final uniform-density slice of each grid through \equaref{eq:deltaN}. Then we coarse-grain the lattice curvature perturbation (or $\delta N$) fields by using a normalized Gaussian window function, i.e.
\begin{align}
\mathcal{R}_R(\mathbf{x})=\int d^3y~G_{R}(\mathbf{x}-\mathbf{y})\mathcal{R}(\mathbf{y}),\qquad G_{R}(\mathbf{r})=\frac{e^{-r^2/(2R^2)}}{(2\pi R^2)^{3/2}}.
\label{eq:gaussian}
\end{align}
We subtract its spatial mean $\bar{\mathcal{R}}=\langle \mathcal{R}_R\rangle$ and define
\begin{equation}
\label{eq:definenu}
\nu(\mathbf{x})
=
\frac{\mathcal{R}_R(\mathbf{x})- \bar{\mathcal{R}}}
{\sigma_{\mathcal R}},
\qquad
\sigma_{\mathcal R}^2
=
\left\langle
(\mathcal{R}_R(\mathbf{x})
- \bar{\mathcal{R}})^2\right\rangle,
\qquad
S_3=\langle\nu^3(\mathbf{x})\rangle,
\qquad
K_{\mathrm{exc}}=\langle\nu^4(\mathbf{x})\rangle-3.
\end{equation}
The one-point probability density is estimated from a normalized histogram of the lattice values of $\nu$. 

\section{Warm-up: Starobinsky inflation}
\label{sec:starobinsky}

We first consider Starobinsky inflation as a benchmark.
The smooth Starobinsky potential produces a nearly scale-invariant curvature power spectrum, providing a benchmark for testing repeated zoom-ins against linear predictions.
The potential in the Einstein frame is \cite{Starobinsky:1980te, Whitt:1984pd}
\begin{equation}
    V(\phi)=\frac{3m^2}{4}
    \left(1-e^{-\sqrt{2/3}\phi}\right)^2,
    \qquad m=1.3\times10^{-5}.
\end{equation}
% The potential approaches a plateau, $V\rightarrow3m^2/4$, at large
% positive $\phi$ and has a quadratic minimum, $V\simeq m^2\phi^2/2$, near the origin.
We initialize the mean field at $\bar\phi=5.5$ on the slow-roll trajectory, where we set the e-fold counts to $N=0$. As the field rolls towards the potential minimum, the increasing slow-roll parameter produces a smoothly red-tilted spectrum. We perform two simulations with $N_{\mathrm{pts}}=64$ and $128$, both evolved to $N=50$. Their parameters are summarized in \tabref{tab:starobinsky_setup}. For both runs, we first evolve eight randomly selected child branches from the first parent simulation. Then, we randomly select one child lattice per branch for the proceeding zoom-ins. For these runs, we use the sparse-integer inheritance prescription described in \secref{sec:zoom_integer}. 

\begin{table}[t]
\centering
\begin{tabular}{lcc}
\hline
$N_{\mathrm{pts}}$ & $64$ & $128$ \\
\hline
Zoom-in factor $A$ & $4$ & $8$ \\
Number of zoom-ins & $30$ & $20$ \\
$a_0L_0H_{\mathrm{init}}$ & $1$ & $2$ \\
$b_{\mathrm{pre}}$ & $8$ & $16$ \\
$b_{\mathrm{post}}$ & $2$ & $2$ \\
Spatial finite-difference order & $6$      & $6$ \\
\hline
\end{tabular}
\caption{Parameters for the Starobinsky benchmark. $H_{\mathrm{init}}$ is the homogeneous initial Hubble parameter used to specify the initial box length.}
\label{tab:starobinsky_setup}
\end{table}

\begin{figure}[t]
    \centering
    \includegraphics[width=0.8\linewidth]{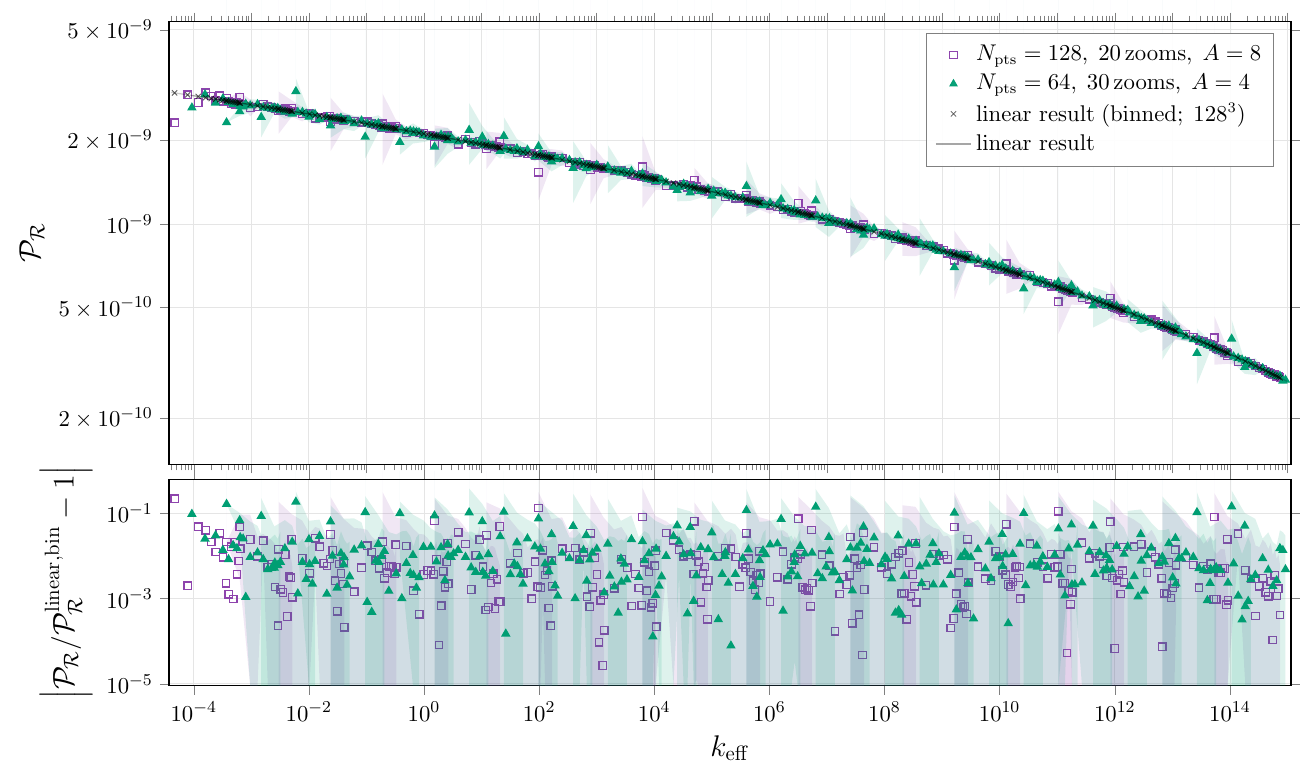}
    
    \caption{
    % \ac{I think here and in fig 8 the binned linear result is too hard to see on the plot, we should find a way to make those crosses more visible}
    Curvature power spectra of Starobinsky inflation simulations with sparse-integer inheritance and periodic boundary conditions. Purple squares denote the $N_{\mathrm{pts}}=128$ run with 20 zooms at $A=8$, and green triangles denote the $N_{\mathrm{pts}}=64$ run with 30 zooms at $A=4$. The upper panel shows $\mathcal{P}_{\mathcal R}$ as a function of the effective comoving wavenumber $k_{\mathrm{eff}}$ of the sixth-order finite-difference stencil. The gray curve shows the unbinned linear prediction, and the black crosses show the linear prediction binned for the $128^3$ run. The lower panel shows the absolute fractional deviation, $\left|\mathcal{P}_{\mathcal R}/ \mathcal{P}_{\mathcal R}^{\mathrm{linear,bin}}-1\right|$. Shaded bands show $1\sigma$ deviation across the eight child patches in each generation. }
\label{fig:starobinsky_spectrum}

\end{figure}

\figref{fig:starobinsky_spectrum} shows the resulting comoving curvature power spectra. We compare the spectra obtained from the zoom-in simulations with the binned linear theory prediction obtained by solving~\equaref{eq:mseq} and binning it on the same bins as the lattice. The bottom panel shows the absolute fractional deviation of each lattice spectrum from the linear prediction binned for that run. For bins having a dimensionless wavenumber $|\mathbf q|\geq4$, the median absolute relative differences are approximately $0.71\%$ and $0.40\%$ for the $64^3$ and $128^3$ runs, respectively. This shows that the zoom-in simulations correctly reproduce the linear-theory result.

\begin{figure}[t]
    \centering
    \includegraphics[width=0.8\linewidth]{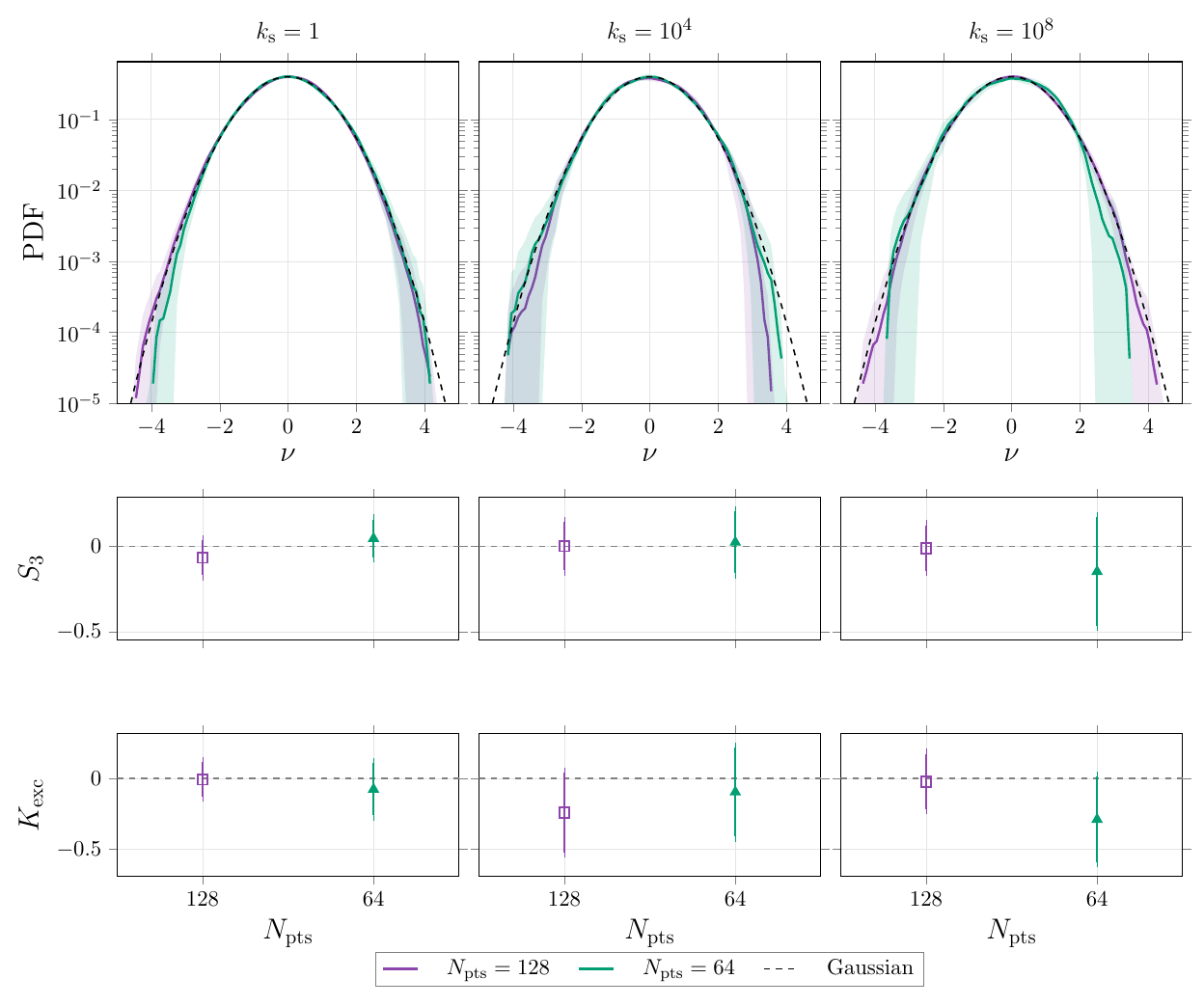}
    \caption{One-point statistics of the coarse-grained $\delta N$ field for the $N_{\mathrm{pts}}=128$ (purple) and $N_{\mathrm{pts}}=64$ (green) simulations of Starobinsky inflation. The upper panel shows the PDFs of $\nu$ as defined in \equaref{eq:definenu}, together with a standard Gaussian (dashed). The middle and lower panels show $S_3=\langle \nu^3\rangle$ and $K_{\mathrm{exc}}=\langle \nu^4\rangle-3$.
    The columns correspond to the comoving smoothing wavenumbers $k_{\mathrm{s}}=1,10^4,10^8$, with $R=k_{\mathrm{s}}^{-1}$. We use the $\delta N$ formalism with a coarse-graining scale described in \secref{sec:deltaN}. For each child, we subtract the spatial mean of the coarse-grained field and divide by its standard deviation before pooling the normalized lattice values with equal weight. The upper panels show the resulting PDFs of $\nu$ defined in \secref{sec:deltaN}. The dashed curve is a standard Gaussian. The middle and lower panels show the skewness and excess kurtosis. Shaded bands and error bars show the $1\sigma$ scatter among the eight children.}
    \label{fig:starobinsky_pdf}

\end{figure}

We also calculate the nonlinear $\delta N$ one-point statistics and show the results in \figref{fig:starobinsky_pdf}. The one-point PDF is close to Gaussian, while the measured skewness and excess kurtosis are also compatible with 0, as expected for this potential.

\section{Oscillatory feature}
\label{sec:oscillatory_starobinsky}

We next test whether the zoom-in scheme can recover a localized feature in the curvature power spectrum. To this end, we add an oscillatory modulation of the form considered in Ref.~\cite{Caravano:2024moy} to the Starobinsky potential. We write
\begin{equation}
    V(\phi) = \frac{3m^2}{4}\left(1-e^{-\sqrt{2/3}\,\phi}\right)^2
    + \Lambda^4\mathcal{W}(\phi)\bigl[\cos u-1\bigr],
    \qquad u=\frac{\phi_{\mathrm{entry}}-\phi}{f},
    \label{eq:oscillatory_starobinsky_potential}
\end{equation}
where $\mathcal{W}(\phi)=[1+\tanh u][1+\tanh(\Delta\phi/f-u)]/4$. 
We use $m=6.6769\times10^{-7}$, $\Lambda^4=1.2684\times10^{-19}$, $f=1.4142\times10^{-4}$, $\Delta\phi=3.0123\times10^{-3}$, and $\phi_{\mathrm{entry}}=8.63747$ to match the dimensionless parameters of the perturbative benchmark (Setup~1 of Ref.~\cite{Caravano:2024tlp}). 
The oscillations transiently amplify a band of subhorizon modes as the field traverses the oscillatory feature. We define $N_{\mathrm{entry}}$ as the time when the first lattice parent reaches the mean-field value $\langle\phi\rangle=\phi_{\mathrm{entry}}$. We denote the scale crossing the horizon at this time as $k_{\mathrm{entry}}=a(N_{\mathrm{entry}})H(N_{\mathrm{entry}})$, which serves as a reference scale when comparing scales across different runs.

For this setup, we run a $512^3$ simulation without zoom-ins as a reference and call this run 512-0. We also run two $256^3$ simulations with one and two zoom-ins at $A=4$, both using the linear-environment prescription. We call these runs 256-1 and 256-2, respectively. Finally, we run a $64^3$ simulation with six zoom-ins at $A=4$ using the cheaper real-space interpolation method introduced at the end of~\secref{sec:comparison}. We call this run 64-6. The parameters for all four runs are summarized in \tabref{tab:oscillatory_starobinsky_setup}. For all zoom-in runs, we use the non-periodic estimator described in \appref{app:estimator} to extract the power spectrum from child lattices.

\begin{figure}[t]
    \centering
    \includegraphics[width=0.8\linewidth]{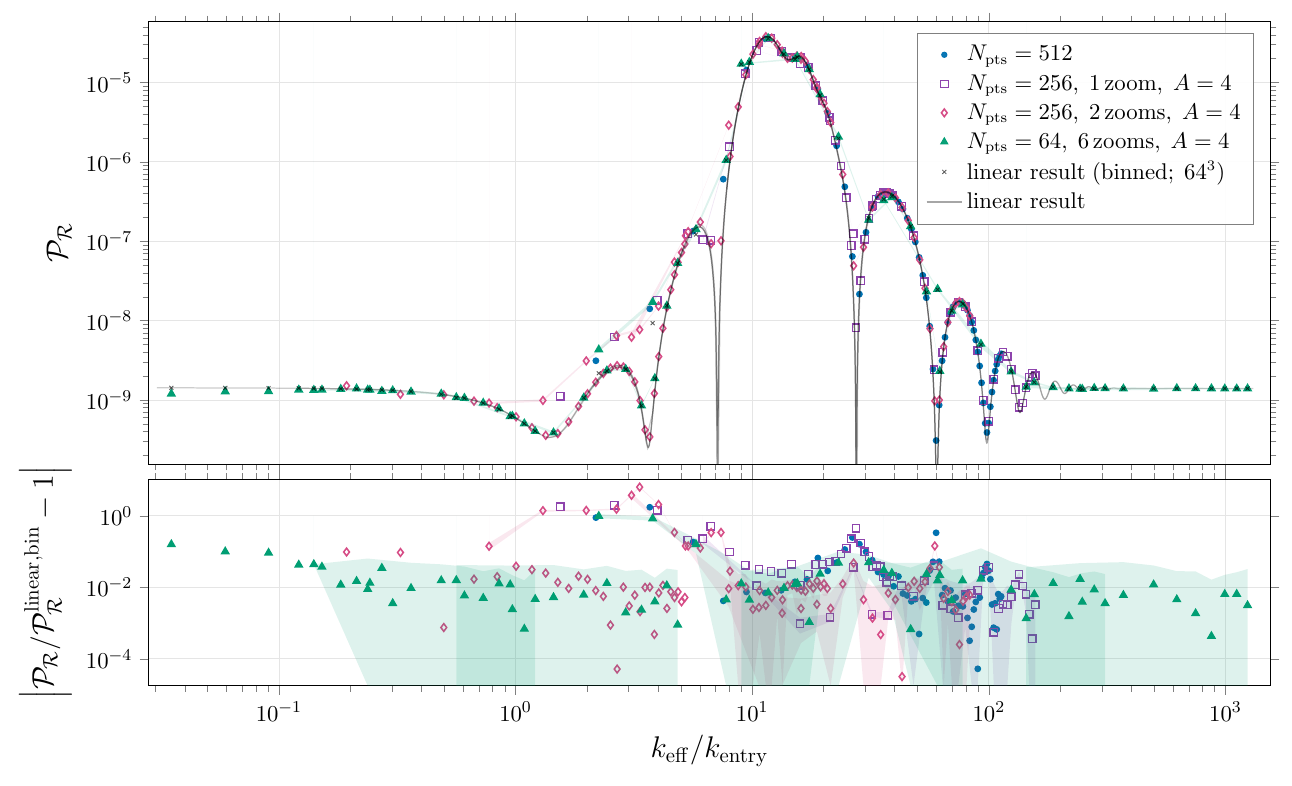}
    \caption{Curvature power spectra of the simulations for the Starobinsky potential with oscillatory features. The upper panel compares the four lattice runs with the unbinned linear prediction (gray curve) and the linear prediction binned for the $64^3$ run (black crosses).
    The lower panel shows $\left|\mathcal{P}_{\mathcal R}/\mathcal{P}_{\mathcal R}^{\mathrm{linear,bin}}-1\right|$. Shaded bands show $1\sigma$ deviation across the eight child patches in each generation.}
    \label{fig:oscillatory_starobinsky_power}
\end{figure}

The spectra in \figref{fig:oscillatory_starobinsky_power} are evaluated approximately 10 e-folds after the time of crossing $N_{\rm entry}$. The four runs recover the peak at $k_{\mathrm{eff}}/k_{\mathrm{entry}}\simeq12$, with $\mathcal{P}_{\mathcal R}\simeq3.6\times10^{-5}$, and follow the subsequent oscillations over their common range. The 64-6 run extends the plotted range to $3.5\times10^{-2}\lesssim k_{\mathrm{eff}}/k_{\mathrm{entry}}\lesssim1.2\times10^3$, compared with approximately $2.2$ to $113$ for the 512-0 run. 

The power spectra from these simulations broadly agree with the linear prediction, except for a localized spot near the first dip, where the simulations show an excess of power compared to the linear theory. The excess is also present in the 512-0 run, so it cannot be attributed to the zoom-in artifacts but it reflects genuine mode coupling, where resonantly amplified short-wavelength modes can source a long-wavelength perturbation. This effect was also reported using lattice simulations in \cite{Caravano:2024tlp}. In the 64-6 run, the second zoom-in lattices stay close to the linear prediction, whereas the third zoom-in lattices show excess power over the same range of wavenumbers, and the same happens for the root lattice and the first zoom-in lattices of the 256-2 run. This difference can be explained by the fact that the second generation zoom-in lattices do not resolve enough UV modes to see this nonlinear effect. One-loop calculations for a similar oscillatory feature also find departures from linear theory \cite{Inomata:2022yte} (see Ref.~\cite{Braglia:2026fzp} for a recent comparison using a mean-field approximation).

\begin{table}[t]
    \centering
    \begin{tabular}{lcccc}
        \hline
        $N_{\mathrm{pts}}$          & $512$(blue)    & $256$(purple)      & $256$(pink)     & $64$(green) \\
        \hline
        Zoom-in factor $A$          & --       & $4$      & $4$      & $4$ \\
        Number of zoom-ins             & $0$      & $1$      & $2$      & $6$ \\
        $a_0 L_0 H_{\mathrm{init}}$ & $0.8774$ & $1.2408$ & $9.9263$ & $1$ \\
        $b_{\mathrm{pre}}$          & --       & $10$     & $10$     & $10$ \\
        $b_{\mathrm{post}}$         & --       & $2.5$    & $2.5$    & $2.5$ \\
        Spatial finite-difference order
                                    & $2$      & $2$      & $6$      & $6$ \\
        \hline
    \end{tabular}
    \caption{Parameters for the four oscillatory Starobinsky
    simulations. }
    \label{tab:oscillatory_starobinsky_setup}
\end{table}

% For the periodic root grids we use the direct Fourier estimator; for the refined grids we combine the inherited and newly injected contributions using the same mode weights in the lattice and linear estimates. Matching these weights is important near the sharp dips, where the bin average need not agree with the unbinned value at the bin center. The linear reference is not convolved with the crop window, so the residuals in the lower panel cannot be attributed to nonlinear evolution alone.

To test the shape of the distribution beyond its power spectrum, we construct the coarse-grained $\delta N$ field on designated comoving scales $k_{\mathrm{s}}/k_{\mathrm{entry}}=5,10, 50,100$ shared by all simulations from the total field and velocity as described in \secref{sec:deltaN}.
For the zoom-in simulations, for each comoving smoothing radius $R=k_{\mathrm{s}}^{-1}$, we select a corresponding generation of the child lattices in each run, balancing resolution of the smoothing scale against the available sampling volume. For example, at $k_{\mathrm s}/k_{\mathrm{entry}}=10$, we use generation~1 of the 256-1 run,
generation~2 of the 256-2 run, and
generation~4 of the 64-6 run. 
For non-periodic child lattices, we truncate the Gaussian kernel at $4R$ along each coordinate and renormalize its weights to unity.
We retain only the smoothed values at points whose entire smoothing kernel lies inside the child lattice, excluding a boundary layer of $4R$, which also guarantees that the smoothing window never crosses a child boundary. For a child box of side length $L_c$, we retain only the smoothed values at points within $[4R,L_c-4R]^3$.
\begin{align}
\mathcal R_R(\mathbf{x})=
\int_{\mathbf{x}+ [-4R,4R]^3}\mathrm d^3y\,
G_R^{\mathrm{tr}}(\mathbf x-\mathbf y)
\mathcal R(\mathbf y),
\qquad
\mathbf x\in[4R,L_c-4R]^3
\end{align}
with a renormalized truncated Gaussian window function
\begin{align}
    G_R^{\mathrm{tr}}(\mathbf r)=G_R(\mathbf r)\frac{1}{\int_{[-4R,4R]^3} dl^3 G_R(\mathbf{l})} \qquad
\mathbf r\in[-4R,4R]^3
\end{align}
where $\int_{\mathbf{x}+[-4R,4R]^3}~dl^3=\int_{x_x-4R}^{x_x+4R}~dl_x\int_{x_y-4R}^{x_y+4R}~dl_y\int_{x_z-4R}^{x_z+4R}~dl_z$ denotes integation over
a cube of side length $8R$ centered at $\mathbf x$. Using the coarse-grained $\mathcal{R}_R$, we compute the one-point PDF and skewness according to \secref{sec:deltaN}.

\begin{figure}[t]
    \centering
    \includegraphics[width=0.8\linewidth]{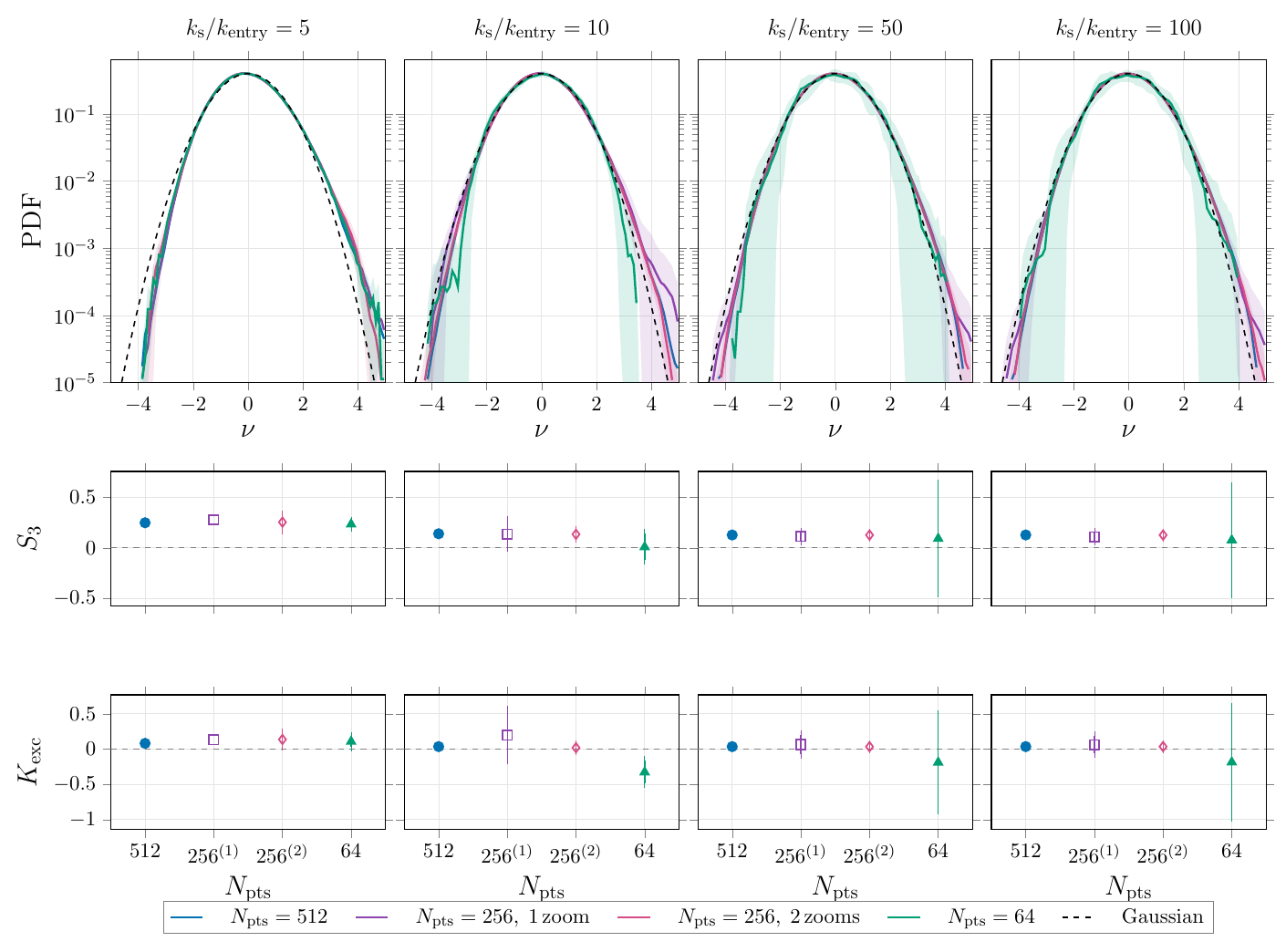}
    \caption{One-point statistics of the coarse-grained $\delta N$ field. The columns correspond to four coarse-graining scales $k_{\mathrm{s}}/k_{\mathrm{entry}}=5,10,50,100$. The upper panel shows the PDFs of $\nu$ as defined in \equaref{eq:definenu}, together with a standard Gaussian (dashed). The middle and lower panels show $S_3=\langle \nu^3\rangle$ and $K_{\mathrm{exc}}=\langle \nu^4\rangle-3$. Shaded bands and error bars show the sample standard deviations across eight children.}
    \label{fig:oscillatory_starobinsky_deltaN}
\end{figure}

\figref{fig:oscillatory_starobinsky_deltaN} and \tabref{tab:coarse_grained_skewness} show positive skewness in all four runs at $k_{\mathrm{s}}/k_{\mathrm{entry}}=5$, with $S_3\simeq0.23-0.28$. Across the 512-0, 256-1, and 256-2 runs, the central values of the skewness range from $0.13$ to $0.14$ at $k_{\mathrm{s}}/k_{\mathrm{entry}}=10$ and from $0.11$ to $0.13$ at $50$ and $100$. The 64-6 run shows substantial scatter among child lattices at these three scales. The positive skewness and its variation with smoothing scale show that the curvature perturbations have scale-dependent non-Gaussianity. This is expected in this model, where the oscillatory feature amplifies a limited band of modes, so the contribution of their nonlinear coupling depends on which scales the smoothing window retains. 

\begin{table}[htbp]
    \centering
  \begin{tabular}{ccccc}
    \hline
    $k_{\mathrm{s}}/k_{\mathrm{entry}}$
        & $N_{\mathrm{pts}}=512$
        & \shortstack{$N_{\mathrm{pts}}=256$(purple)}
        & \shortstack{$N_{\mathrm{pts}}=256$(pink)}
        & $N_{\mathrm{pts}}=64$ \\
    \hline
    $5$   & $0.2474$ & $0.2761$           & $0.2532\pm0.0694$ & $0.2323\pm0.0293$ \\
    $10$  & $0.1388$ & $0.1362\pm0.1293$ & $0.1337\pm0.0353$ & $0.0085\pm0.1297$ \\
    $50$  & $0.1269$ & $0.1126\pm0.0394$ & $0.1262\pm0.0143$ & $0.0917\pm0.5338$ \\
    $100$ & $0.1269$ & $0.1094\pm0.0393$ & $0.1259\pm0.0140$ & $0.0762\pm0.5222$ \\
    \hline
\end{tabular}
    \caption{Skewness $S_3$ of the $\delta N$ field at four coarse-graining scales.  Uncertainties are sample standard deviations across eight child grids.}
    \label{tab:coarse_grained_skewness}
\end{table}

The skewness of the coarse-grained curvature perturbation receives contributions from bispectra of different sizes and shapes. To estimate the equilateral component, we measure the equilateral bispectrum in the 512-0 and 256-2 simulations. We estimate the bispectrum by averaging products of three Fourier coefficients over closed wavevector triangles. For the equilateral configuration, all three wavevectors lie in the same shell. To suppress boundary leakage, before doing Fourier transformation, we apply a cosine taper using the window function $W(x,y,z)=w(x)w(y)w(z)$ to the $\delta N$ field. Here $w(x)=1$ in the central interval $0.12L\leq x\leq0.88L$ and decreases smoothly to zero
across a layer of width $0.12L$ to the boundary following $w(x)=1/2[1-\cos(\pi \min(x,L-x)/(0.12L))]$. This numerical treatment suppresses boundary-induced spectral leakage. Thus the bispectrum of equilateral shape is calculated using 
\begin{align}
\mathcal{B}^{\mathrm{eq}}_{\mathcal{R}}=\frac{V^2}{(2\pi)^4\langle W^3\rangle}\left\langle [k_{\rm{eff}}(\mathbf{k}_1) k_{\rm{eff}}(\mathbf{k}_2) k_{\rm{eff}}(\mathbf{k}_3)]^2\rm{Re}[\mathcal{R}_{\mathbf{k}_1}^W\mathcal{R}_{\mathbf{k}_2}^W\mathcal{R}_{\mathbf{k}_3}^W]\right\rangle_{\mathrm{eq,shell}},
\end{align}
where the average is taken over closed triangles satisfying $\mathbf k_1+\mathbf k_2+\mathbf k_3=\mathbf 0$, with all three wavevectors lying in the same shell $b$. Here, as described before, ${\mathcal{R}_\mathbf{k}^W}$ represents the Fourier coefficient after applying the window function to the real space $\delta N$ field, $\mathcal{R}^W=W(\mathcal R-\langle{W \mathcal{R}}\rangle/\langle{W}\rangle)$. Here $V=L^3$ is the comoving box volume, $\langle W^3\rangle$ is the spatial average of the cubed taper, and the Fourier coefficients use the normalization of \secref{sec:statistics}.

We also check the robustness of the treatment with a wider taper of $0.20L$ at each boundary and find that the results remain stable.

\begin{figure}[t]
    \centering
    \includegraphics[width=0.8\linewidth]{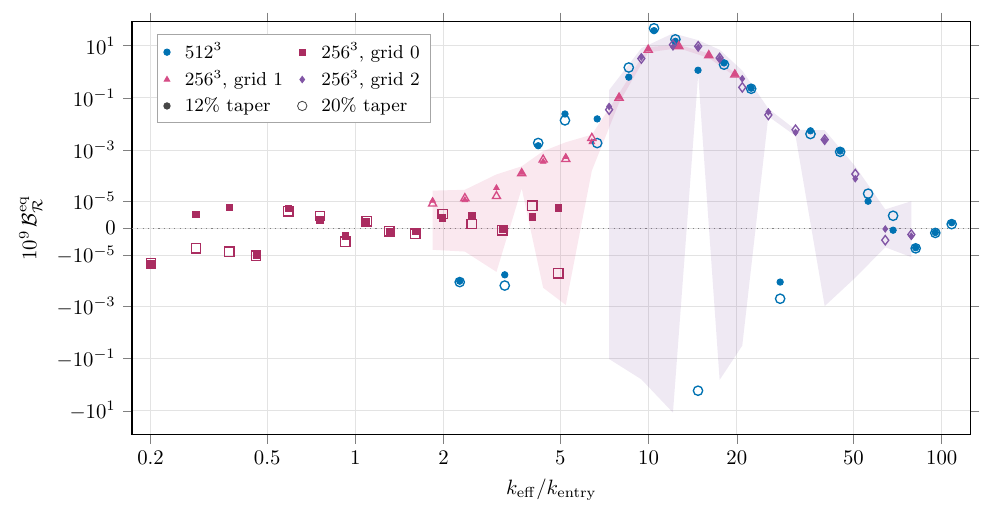}
    \caption{Equilateral bispectrum of the $\delta N$ field on a common final uniform-density slice in the Starobinsky model with oscillatory features. We compare the 512-0 run with the root (grid~0) and two refinement levels (grids~1 and~2) of the 256-2 zoom-in run. All three legs lie in the same Fourier shell for the equilateral configuration. Solid and hollow markers denote taper widths of $0.12L$ and $0.20L$ at each boundary, respectively; shaded bands show the $1\sigma$ scatter among child lattices for the $0.12L$ taper. The vertical axis shows $10^9\mathcal{B}_{\mathcal{R}}^{\mathrm{eq}}$.}
    \label{fig:oscillatory_starobinsky_bispectrum}
\end{figure}

As shown in \figref{fig:oscillatory_starobinsky_bispectrum}, both runs show a positive enhancement in the equilateral bispectrum around $k_{\mathrm{eff}}/k_{\mathrm{entry}}\simeq10\sim 15$, near the peak of the power spectrum. 

\section{Conclusions and discussion}

We have developed a zoom-in scheme for lattice simulations of inflation. At each zoom-in time, we transfer the field and its velocity to smaller child boxes and initialize newly resolved short-wavelength modes. This shifts the resolved band towards larger comoving wavenumbers without increasing the number of lattice points per box. Successive zoom-ins thus allow us to follow modes around horizon crossing over a much wider range of scales than a single fixed grid with the same number of lattice points can resolve.

The inherited field on a child grid is generally non-periodic, so imposing periodic boundary conditions on the full field can introduce artifacts. We develop three methods to address this problem. Sparse-integer inheritance keeps only modes that are periodic on the child grid and rescales them to preserve their volume-normalized power, and neglects the other modes. The linear-environment and parent-matched prescriptions retain the non-periodic modes, accounting for the parent's influence through the child boundary conditions. The linear-environment method evolves these modes with a linear transfer matrix. The parent-matched method obtains the boundary values from the parent grids, which evolve alongside the child. Both methods closely reproduced the field and velocity of the cropped fine-grid reference in the two-e-fold test.

For Starobinsky inflation, we evolved the $64^3$ and $128^3$ simulations over 50 e-folds. For bins centered at $|\mathbf q|\geq 4$, the median absolute deviations from the binned linear curvature spectrum were $0.71\%$ and $0.40\%$. The $\delta N$ distributions were close to Gaussian. In the Starobinsky model with an oscillatory feature, the zoom-in runs recovered the main peak and the subsequent oscillations seen in the fixed-grid $512^3$ simulation, while the six-zoom-in $64^3$ run extended the spectrum towards both the infrared and the ultraviolet. A localized excess over linear theory near the first dip also appears in the $512^3$ fixed-grid run, consistent with nonlinear mode coupling. This result is compatible with earlier lattice simulations \cite{Caravano:2024tlp} and analytical calculations \cite{Inomata:2022yte}. The skewness measurements show scale-dependent non-Gaussianity. The 512-0 and 256-2 runs both show an enhancement of the equilateral bispectrum near the power-spectrum peak.

The zoom-in scheme developed in this work transfers information from parent grids to child grids. A natural extension would be to include child-to-parent backreaction by feeding the coarse-grained effects of short-wavelength dynamics on the child grid back into the parent evolution. A further extension would be to reconstruct high-resolution curvature maps beyond the sampled child volume, building on the reconstruction of the inflaton field and its velocity across the full parent volume demonstrated in \appref{app:total_map_reconstruction}. These curvature maps could be used to generate initial conditions for high-resolution simulations of structure formation \cite{Jamieson:2026fsw}. Moreover, extending the zoom-in scheme to axion--U(1) inflation \cite{Caravano:2022epk,Figueroa:2024rkr} could increase the dynamical range of lattice simulations and allow nonlinear gauge-field backreaction to be followed over longer intervals, which remains a key challenge in understanding the full inflationary dynamics of this model.

\section{Code and Data availability}

The simulation code and data supporting the results presented in this work will be made publicly available upon journal publication.

\section*{Acknowledgments}
X.W. thanks Elisa Ferreira,  Masahiro Takada and Vincent Vennin for useful discussions.
This work is supported by Forefront Physics and Mathematics Program to Drive Transformation (FoPM), a World-leading Innovative Graduate Study (WINGS) Program, the University of Tokyo [X.W.], by JSPS KAKENHI Numbers JP26KJ0757 [X.W.],  JP26K17136 [L.T.],  JP24K00624 [M.S.], by JST PRESTO JPMJPR26287514 [L.T.], by a Royal Society International Collaboration Award Round 2 Japan 252140 [L.T.], by a CD3 Google seed grant [X.W.], and by the European Union’s Horizon Europe research and innovation programme under Marie Sk\l{}odowska-Curie grant agreement No. 101202657 [A.C.].
Kavli IPMU is supported by the World Premier International Research Center Initiative (WPI), MEXT, Japan. X.W. also acknowledges support by grant 63132 from the John Templeton Foundation, as recipient of an Enrico Fermi Fellowship awarded through the Center for SpaceTime and the Quantum. The opinions expressed in this publication are those of the author(s) and do not necessarily reflect the views of the respective funding body. 

% X.W. acknowledges insightful inputs from the international molecule workshop at YITP International Molecule-type Workshop and APCTP Global Cooperation Workshop “Theoretical Challenges Towards nonlinearities from the Early Universe”, where the work is presented.

\appendix

\section{Initialization and small-scale power injection}
\label{app:initialization}
The initialization supplies both the field perturbation and its velocity in \equaref{eq:zoom_cauchy}. 
% Rather than assigning an instantaneous vacuum
% at every zoom, we evolve the linear mode functions to the time at which
% the modes are introduced. A mode that has already approached the horizon
% or passed through a feature therefore retains the evolution predicted by
% \equaref{eq:mseq}. 
For the field mode function $Q_{\mathbf k}$, the adiabatic frequency is
\cite{Bunch:1978yq}
\begin{equation}
 \omega_{\mathbf k}^2
 =k_{\mathrm{eff}}^2
  +a^2\left[V_{\phi\phi}(\bar\phi)+\Delta m_g^2\right]
  -\frac{a''}{a},
 \label{eq:wkb_omega}
\end{equation}
where primes denote conformal time derivatives. The background and metric correction are those used in the linear perturbation equation, and $k_{\mathrm{eff}}$ is evaluated with the spacing and difference order of the grid on which the mode will be initialized.

For each mode, we choose a starting time $N_{*,\mathbf k}$ no later than its initialization time $N_{\mathrm b}$. At this starting time the mode must be sufficiently subhorizon, with positive, slowly varying $\omega_{\mathbf k}^2$, so that the adiabatic initial conditions are applicable:
\begin{equation}
 Q_*=\frac{1}{a_*\sqrt{2\omega_*}},
 \qquad (\partial_NQ)_*=
 \left[-1-\frac12\partial_N\ln\omega_*
       -i\frac{\omega_*}{a_*\bar H_*}\right]Q_*.
\end{equation}
The star indicates evaluation at $N_{*,\mathbf k}$. 
The adiabatic conditions need not hold at $N_{\mathrm b}$: the mode
function and its derivative can be propagated to that time using
\equaref{eq:mseq},
\begin{equation}
 \begin{pmatrix}Q\\ \partial_NQ\end{pmatrix}_{N_{\mathrm b}}
 =\mathbf T_{\mathbf k}(N_{\mathrm b},N_{*,\mathbf k})
 \begin{pmatrix}Q\\ \partial_NQ\end{pmatrix}_{N_{*,\mathbf k}}
\end{equation}
if necessary.
Writing the linear mode equation as a first-order system for the mode function and its e-fold derivative gives the transfer matrix used in the linear-environment method:
\begin{equation}
\mathbf T_{\mathbf k}(N_{\mathrm b},N_*)
=\mathcal T\exp\!\left[
\int_{N_*}^{N_{\mathrm b}}\!\dd\widetilde N\,
\begin{pmatrix}
0 & 1\\[2pt]
-\dfrac{k_{\mathrm{eff}}^2}{a^2\bar H^2}
-\dfrac{V_{\phi\phi}(\bar\phi)+\Delta m_g^2}{\bar H^2}
&
-\left(3+\partial_{\widetilde N}\ln\bar H\right)
\end{pmatrix}
\right],
\label{eq:mode_transfer_matrix}
\end{equation}
with all background quantities evaluated at $\widetilde N$. Time ordering, denoted by $\mathcal T$, places the later evolution to the left of the earlier evolution. In practice, we integrate the mode equation with RK4.

Using the normalized Fourier coefficients of \secref{sec:statistics}, the random field and velocity at $N_{\mathrm b}$ are constructed as follows. For one member of each nonzero pair
$(\mathbf k,-\mathbf k)$, take two independent complex Gaussian variables,
\begin{equation}
 \xi_{\mathbf k}^{(\ell)}
 =\sqrt{-\ln U_1^{(\ell)}}\,e^{2\pi iU_2^{(\ell)}},
 \qquad \ell=1,2,
\end{equation}
where the four $U$ variables are independent and uniform on $(0,1)$. The Fourier coefficients are
\begin{equation}
 \begin{pmatrix}
{\delta\phi}_{\mathbf k}\\
 {\delta(\partial_N\phi)}_{\mathbf k}
 \end{pmatrix}
 =
 \frac{1}{L^{3/2}}
 \left.
 |Q_{\mathbf k}|
 \begin{pmatrix}
 \xi_{\mathbf k}^{(1)}\\[4pt]
 \operatorname{Re}\!\left(
     \dfrac{\partial_NQ_{\mathbf k}}{Q_{\mathbf k}}
 \right)\xi_{\mathbf k}^{(1)}
 +
 \operatorname{Im}\!\left(
     \dfrac{\partial_NQ_{\mathbf k}}{Q_{\mathbf k}}
 \right)\xi_{\mathbf k}^{(2)}
 \end{pmatrix}
 \right|_{N_{\mathrm b}}.
 \label{eq:fft_mode_norm}
\end{equation}
% The factor $L^{-3/2}$ converts the continuum mode-function normalization to the finite-volume Fourier convention. 
The two random variables specify the joint Gaussian field and velocity fluctuations, including
their correlation. The remaining coefficients follow from Hermitian symmetry,
\begin{equation}
 {\delta\phi}_{-\mathbf k}
 ={\delta\phi}_{\mathbf k}^{*},
 \qquad
 {\delta(\partial_N\phi)}_{-\mathbf k}
 ={\delta(\partial_N\phi)}_{\mathbf k}^{*},
\end{equation}
so that real-space fields are real.

On grid 0, the inverse transforms of the Fourier components are added to the homogeneous initial field and velocity. At a zoom-in, this sampling is applied only to the new modes allowed by the injection cutoff. The linear-environment method uses the same transfer equation, but its initial coefficients come from the evolved parent periodic component. Coefficients stored from previous generations continue from their values at the zoom-in time.

\section{Power spectrum estimator for non-periodic grids}
\label{app:estimator}

The decomposition in \equaref{eq:zoom_environment_split} also allows us to estimate the spectrum without Fourier transforming the fractional modes on the child box. At the analysis time, we subtract the same $B_1$ that supplies the non-periodic component of the boundary and Fourier transform $r_1$. Notice that changes generated by the nonlinear child evolution remain in $r_1$.
The subtracted contribution $B_1$ is not simply discarded; instead, the power carried by $B_1$ is then included separately. We assign the contribution of each fractional mode to a bin according to its effective wavenumber $k_{\mathrm{eff}}$ evaluated with the child spacing. In this way a fractional mode contributes at its own wavenumber instead of spreading across the child FFT bins.

For one parent-child pair, the resulting power spectrum estimate in bin $b$ is
\begin{equation}
 {\mathcal P}_{\delta\phi,b}
 =\frac{1}{2\pi^2}
  \sum_{j\in\mathcal V_b}
  \mu_{b,j}\,
  k_{\mathrm{eff}}^3(\mathbf k_j)|a_j|^2,
 \label{eq:estimator}
\end{equation}
where $\mathcal V_b$ includes the following three categories of contributions, with different weights listed below. 
\begin{center}
\begin{tabular}{lcc}
\hline
Mode & Coefficient $a_j$ & Weight $\mu_{b,j}$ \\
\hline
Periodic inherited ($\mathbf q_0\in A\mathbb Z^3$)& $r_{q,\mathbf q_1}$& $\alpha_b V_0/N_{0,b}$\\
Non-periodic inherited ($\mathbf q_0\notin A\mathbb Z^3$)& $c_{0,\mathbf q_0}$& $\alpha_b V_0/N_{0,b}$\\
Newly initialized
 & $r_{1,\mathbf q_1}$& $V_1/N_{1,b}$\\
\hline
\end{tabular}
\end{center}

% Here $r_{c,\mathbf q_c}$ is measured from the current child periodic component, whereas $c_{p,\mathbf q_p}$ is the coefficient used in
% $B_c$ at the same time. 
Here $V_0=L_0^3$ and $V_1=L_1^3$ are the parent and child comoving volumes.
Within bin $b$, $N_{i,b}$ and $N_{u,b}$ count the child Fourier vectors assigned to the inherited and newly initialized parts and $N_{1,b}=N_{i,b}+N_{u,b}$. And $N_{0,b}$ includes both the periodic and non-periodic inherited vectors from the parent in that bin. For a nonempty child bin, $\alpha_b=N_{i,b}/N_{1,b}$ is the fraction of child-grid Fourier vectors assigned to inherited modes.
% The last row uses the evolved residual coefficients
% at the newly initialized vectors, rather than their initial vacuum amplitudes.
Therefore, the first two rows must have the same weight since they account for the modes inherited from the parent together in the bin, including non-periodic and periodic.  Multiplying both coefficient sums
by $V_0/N_{0,b}$ puts the retained and restored contributions on the same normalization. The factor $\alpha_b$ then combines this inherited estimate with the newly initialized part, whose weight is
$(1-\alpha_b)V_1/N_{u,b}=V_1/N_{1,b}$ when $N_{u,b}>0$.
For repeated zooms, the same procedure is applied successively to the earlier generations contributing to the bin.

\begin{figure}[t]
    \centering
    \includegraphics[width=1\linewidth]{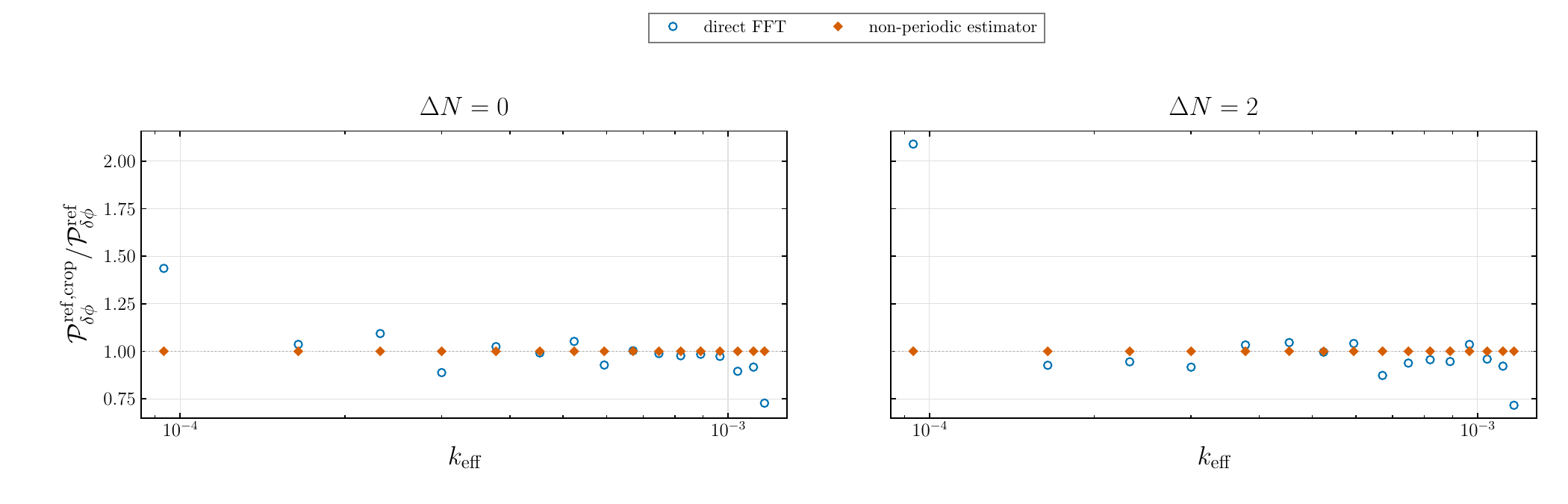}
    \caption{Cropped reference power divided by full-reference power at $\Delta N=0$ and $2$. Direct FFT (blue hollow dots) transforms the
    cropped field, while the non-periodic estimator (orange diamonds)
    combines the retained and fractional contributions using \equaref{eq:estimator}. Agreement with unity verifies recovery of the power spectrum obtained by performing FFT in the reference grid in a designed $k$ bin.}
\label{fig: estimator}
\end{figure}

The estimator \equaref{eq:estimator} therefore avoids crop-window leakage from the known non-periodic component from the parent while retaining child-generated corrections in the periodic component. In the same test run presented in \figref{fig:zoom_boundary_comparison}, 
using the estimator for the cropped reference recovers the power spectrum measured using direct FFT on the full reference grid at both
$\Delta N=0$ and $2$, whereas direct FFT shows deviations caused by spectral leakage. The result is shown in \figref{fig: estimator}.

\section{The dispersion relation}
\label{app:dispersion}

The finite spatial resolution of a lattice modifies the propagation of field perturbations. In a zoom-in simulation this effect requires particular care, since an inherited configuration is represented on grids with different spacings. Here we distinguish the comoving wavevector, which specifies the spatial configuration, from the effective wavenumber, which
determines the action of the discrete Laplacian. We then describe their use in the calculation and the matching of the parent and child spatial operators.
\subsection{The effective lattice wavenumber}
For the spacing $\Delta x$ of \equaref{eq:zoom_geometry}, the discrete
Laplacian acts on a plane wave as
\begin{equation}
 \nabla_{\mathrm{lat}}^2e^{i\mathbf k\cdot\mathbf x_{\mathbf j}}
 =-k_{\mathrm{eff}}^2(\mathbf k;\Delta x)
   e^{i\mathbf k\cdot\mathbf x_{\mathbf j}}.
 \label{eq:disp_definition}
\end{equation}
For example, the second-order centered difference is
\begin{equation}
(\nabla_{\mathrm{lat}}^2\phi)_{\mathbf j}
=
\frac{1}{(\Delta x)^2}
\sum_{\alpha=1}^{3}
\left(
\phi_{\mathbf j+\mathbf e_\alpha}
-2\phi_{\mathbf j}
+\phi_{\mathbf j-\mathbf e_\alpha}
\right),
\label{eq:laplacian_second_order}
\end{equation}
where $\mathbf e_\alpha$ denotes a unit displacement along the $\alpha$-th lattice direction. Substituting a plane wave gives
\begin{equation}
 \frac{e^{ik_\alpha\Delta x}-2+e^{-ik_\alpha\Delta x}}
      {(\Delta x)^2}
 =-\frac{4}{(\Delta x)^2}
   \sin^2\!\left(\frac{k_\alpha\Delta x}{2}\right),
 \label{eq:disp_second}
\end{equation}
which yields the second-order $k_{\mathrm{eff}}$ used in the main text \cite{Caravano:2021pgc}. Its dependence on each component of $\mathbf k$ explains why waves of equal $|\mathbf k|$ can have slightly
different lattice frequencies. The original $k$ is only recovered when $|k_\alpha\Delta x|\ll1$ in every direction. As a result, a mode with same $k$ evolves differently on grids with different spacing, which could cause a systematic mismatch between the child grid and parent grid when representing the same modes.

\subsection{Sixth-order differences and the injected band}
To reduce the error induced by $k_{\rm eff}$, we use the sixth-order centered Laplacian when computing spatial gradients, given by
\begin{equation}
 (\nabla_{\mathrm{lat}}^2 \phi)_{\mathbf j}
 =\frac{1}{(\Delta x)^2}\sum_{\alpha=1}^{3}\sum_{r=1}^{3}c_r
 \left(\phi_{\mathbf j+r\mathbf e_\alpha}
       +\phi_{\mathbf j-r\mathbf e_\alpha}-2\phi_{\mathbf j}\right),
 \qquad
 (c_1,c_2,c_3)=
 \left(\frac32,-\frac3{20},\frac1{90}\right).
 \label{eq:disp_stencil}
\end{equation}
The corresponding effective wavenumber is
\begin{equation}
 k_{\mathrm{eff},6}^2(\mathbf k;\Delta x)
 =\frac{4}{(\Delta x)^2}\sum_{\alpha=1}^{3}
 \sin^2\!\left(\frac{k_\alpha\Delta x}{2}\right)
 \left[
  1+\frac{\sin^2(k_\alpha\Delta x/2)}{3}
   +\frac{8\sin^4(k_\alpha\Delta x/2)}{45}
 \right].
 \label{eq:disp_sixth}
\end{equation}
Expanding at fixed $\mathbf k$ gives
\begin{equation}
 k_{\mathrm{eff},6}^2
 =k^2-\frac{(\Delta x)^6}{560}\sum_{\alpha=1}^{3}k_\alpha^8
   +O\!\left((\Delta x)^8k^{10}\right),
 \qquad k=|\mathbf k|.
 \label{eq:disp_expansion}
\end{equation}
The leading relative error is therefore of order $(k\Delta x)^6$. We therefore restrict the vacuum power injection by
\begin{equation}
 |q^i|<\frac{N_{\mathrm{pts}}}{4}
 \quad(i=x,y,z)
 \qquad\Longrightarrow\qquad
 \left|
 \frac{k_{\mathrm{eff},6}(\mathbf k;\Delta x)}{k}-1
 \right|<0.93\%.
 \label{eq:disp_cut}
\end{equation}

To have the computations and analyses consistent with discretized simulation,  we use $k_{\mathrm{eff}}^2$ instead of $k^2$ in the gradient term of the linear perturbation equations, including the linear transfer calculation and the evolution of the linear-environment boundary conditions. The same dispersion relation also enters the vacuum initialization.  For power spectrum analysis, $k_{\mathrm{eff}}$ is evaluated for each Fourier vector before bin averaging. By contrast, the original comoving wavevector $\mathbf k$ is retained in Fourier phases and parent-child coordinate mappings, since it determines the field configuration in real space.

\section{Reconstruction of the refined field map}
\label{app:total_map_reconstruction}

This appendix describes the reconstruction of a refined periodic parent from one pair of parent and child lattices with the same number of points.
Let the parent and child have side lengths $L_0=A L_1$, where $A$ is an integer, and let each have $N_{\rm pts}^3$ points. We first Fourier refine the parent to a grid with $(A N_{\rm pts})^3$ points covering the same parent box. Its spacing is then equal to the child spacing. We denote this reconstructed refined parent field by $\phi_{\mathrm{rec}}(\mathbf y,N)$.

 Following the zoom-in prescriptions, the child field is decomposed as
\begin{equation}
    \phi_1(\mathbf x,N)=B_1(\mathbf x,N)+r_1(\mathbf x,N).
\end{equation}
Here $B_1$ contains the parent modes that are fractional with respect to the child box, whereas $r_1$ contains the periodic inherited modes, newly initialized modes, and child evolutions. The reconstruction uses the parent to supply the fractional contribution and the child periodic component $r_1$ to supply the integer contribution to the refined larger grid.

For a periodic parent, the fractional contribution at the child positions is
\begin{equation}
    B_1(\mathbf x,N)
    =
    \sum_{\mathbf q_0\notin A\mathbb Z^3}
    c_{0,\mathbf q_0}(N)
    e^{i(2\pi\mathbf q_0/L_0)\cdot
    (\mathbf X_1-\mathbf X_0+\mathbf x)}.
\end{equation}
The coefficients used here must be the same as those supplying the boundary environment at the same time $N$. They are evolved by the linear transfer matrix in the linear-environment prescription, or obtained from the evolved parent in the parent-matched prescription.

After subtracting $B_1$, we Fourier transform the periodic component:
\begin{equation}
    r_1(\mathbf x,N)
    =
    \sum_{\mathbf q_1}
    {r_{1,\mathbf q_1}}(N)
    e^{i(2\pi\mathbf q_1/L_1)\cdot\mathbf x}.
    % \qquad
    % r_{c,\mathbf q_c}
    % =\frac{\widehat r_{c,\mathbf q_c}}{N_{\rm pts}^3}.
\end{equation}
All sums are restricted to the represented Fourier bands. We retain the zero mode of $r_1$, including any change in its mean during evolution, rather than selecting only the vectors initialized at the zoom-in moment.

The field over the refined parent grid is then reconstructed as
\begin{align}
    \phi_{\rm rec}(\mathbf y,N)
    &=
    \sum_{\mathbf q_0\notin A\mathbb Z^3}
    c_{0,\mathbf q_0}(N)
    e^{i(2\pi\mathbf q_0/L_0)\cdot\mathbf y}
    \nonumber\\
    &\quad+
    \sum_{\mathbf q_1}
    r_{1,\mathbf q_1}(N)
    e^{i(2\pi\mathbf q_1/L_1)\cdot
    (\mathbf y+\mathbf X_0-\mathbf X_1)}.
    \label{eq:reconstruction_field}
\end{align}
The first term preserves the fractional parent modes at their original comoving wavevectors. The second periodically extends the child periodic component over the parent box. This means outside the sampled child, this construction assumes a periodic continuation of $r_1$.
Each child index $\mathbf q_1$ corresponds to $\mathbf q_0=A\mathbf q_1$ on the refined parent grid. The position-dependent phase ensures that the periodic component is placed at the original child location. Integer parent modes are not added separately, since their evolved contribution is already included in $r_1$. If the reconstruction involves more than one generation of the child grids, the non-periodic contributions from their original grid where they are periodic must also be retained as described in \equaref{eq:non-periodic_env}.

% When using unnormalized FFT arrays, the corresponding embedding multiplies the residual FFT coefficients by $A^3$, solely to account for the larger number of grid points.

The same reconstruction is applied to $\partial_N\phi$, using the parent velocity coefficients and the Fourier coefficients of $\partial_N r_1$. At the original child lattice points,
\begin{equation}
    \phi_{\rm rec}(\mathbf X_1-\mathbf X_0+\mathbf x,N)
    =\phi_1(\mathbf x,N),
\end{equation}
and similarly for the velocity. This equality follows from the construction and is not an independent validation against the reference.

The power spectrum can also be calculated from the reconstructed total field using an FFT on the refined parent grid, with volume $L_0^3$ and $(A N_{\rm pts})^3$ points. The reconstructed field and velocity can also be used to calculate the one-point statistics and the nonlinear $\delta N$ according to the procedures in the main text.

\begin{figure}[t]
    \centering
        \includegraphics[width=1\linewidth]{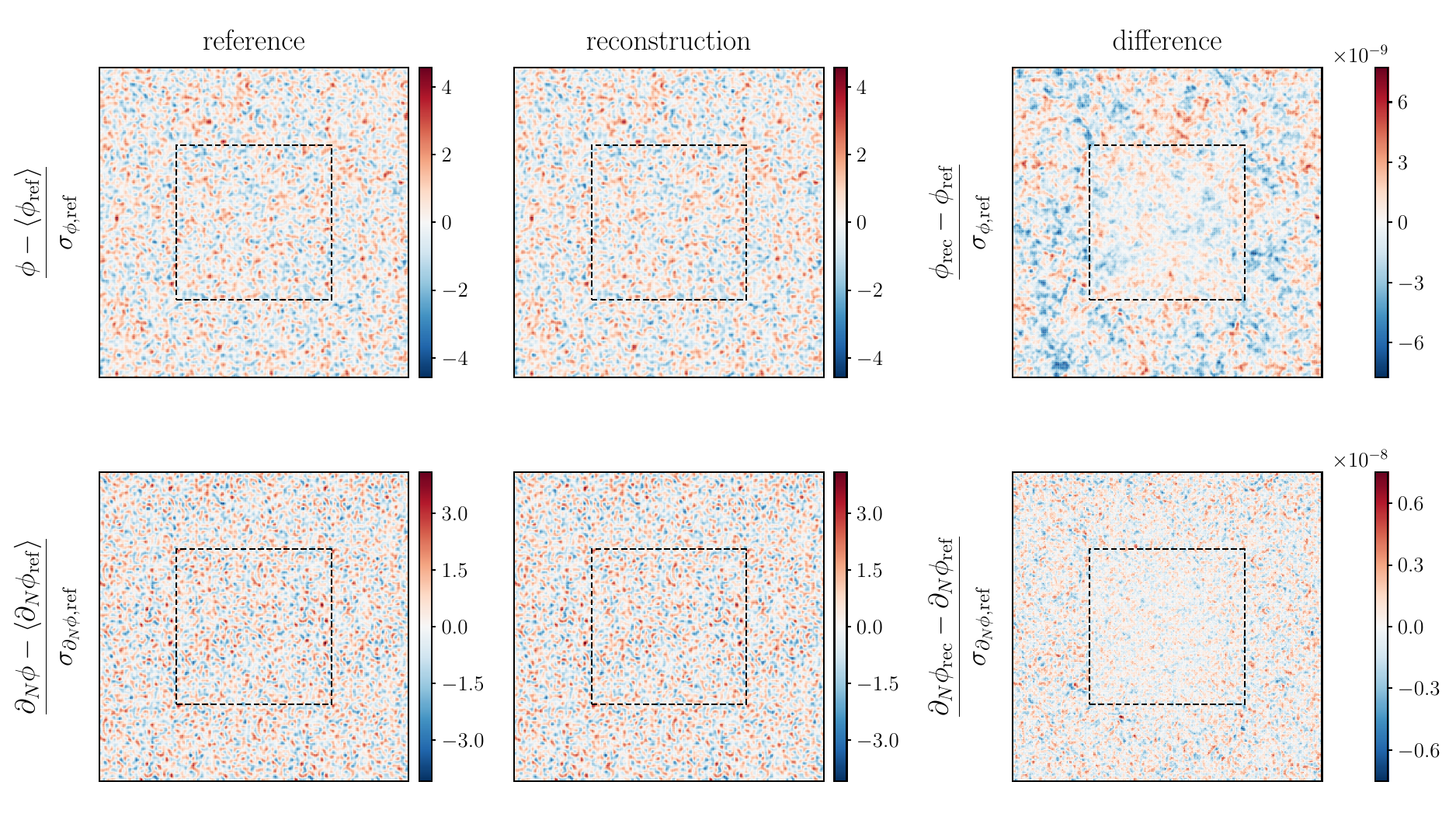}
    \includegraphics[width=0.7\linewidth]{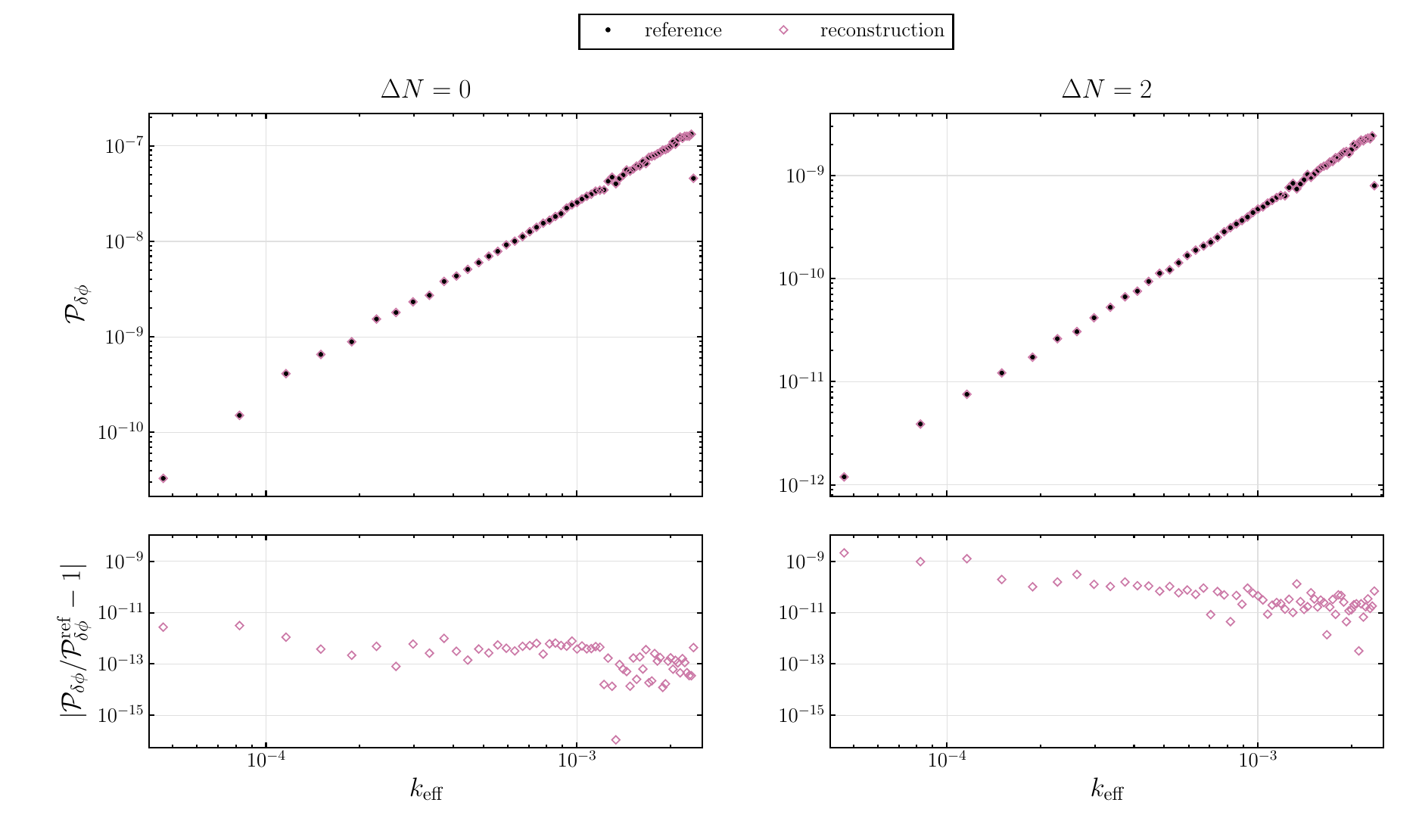}
    \caption{
Reconstruction of the field over the refined parent grid in a two-e-fold Starobinsky simulation with $A=2$. The  parent and child each contain $128^3$ points, while the independently evolved fine reference contains $256^3$ points over the full parent box. \textit{Top:} The two rows show the field $\phi$ and its velocity $\partial_N\phi$ at $\Delta N=2$. The columns show the reference, reconstruction, and their difference. In each row, the first two columns use the reference mean and standard deviation of the quantity shown, and the difference panels are normalized by the same reference standard deviation. Dashed squares mark the original child region. \textit{Bottom:} Field power spectra at $\Delta N=0$ and $2$, shown in the left and right columns. Black markers denote the reference, and pink diamonds denote the reconstruction. Both spectra use the same Fourier vectors and bins obtained from the fine grid. The lower panels show $|\mathcal P_{\delta\phi}/\mathcal P_{\delta\phi}^{\mathrm{ref}}-1|$.
}
\label{fig:reconstruction}

\end{figure}

The reconstruction is limited by the one-way evolution: the small-scale dynamics resolved only by the child do not feed back into the parent. Consequently, agreement with a fully coupled fine-grid simulation is not guaranteed when this feedback becomes important. We leave the implementation of child-to-parent feedback for future work.

We test this reconstruction in a two-e-fold simulation with the Starobinsky potential, taking $m=6\times10^{-6}$ and $\bar\phi_0=5$. The parent and child each contain $128^3$ points, with $L_0=2L_1$, while an independently evolved $256^3$ reference covers the full parent box at the child spacing. The reference is initialized with the same parent modes and the same child modes periodically repeated across the parent volume with a spatial period of the comoving box length of the child. All grids share a common homogeneous background, and the parent initialization and spatial operator use the child-grid dispersion relation.

In \figref{fig:reconstruction}, we compare the reconstructed field map at $\Delta N=2$ directly with the fine reference and show the field power spectra measured by FFT. The reconstructed field has an RMS difference of approximately $1.8\times10^{-9}\sigma_{\phi,\mathrm{ref}}$, while the relative power spectrum difference remains below $2.2\times10^{-9}$.

\bibliographystyle{utphys}
\bibliography{reconstruct}

\end{document}